\documentclass[twocolumn,amsthm]{autart}

\usepackage{graphics} % for pdf, bitmapped graphics files
\usepackage{epstopdf}
\usepackage{amsmath,amssymb,nccmath} % assumes amsmath package installed
\usepackage{algorithm}
\usepackage{algpseudocode}

\newtheorem{theorem}{Theorem}[section]
\newtheorem{corollary}[theorem]{Corollary}
\newtheorem{lemma}[theorem]{Lemma}
\newtheorem{remark}[theorem]{Remark}
\newtheorem{problem}[theorem]{Problem}

\newtheorem{definition}[theorem]{Definition} 
 
\newtheorem{example}[theorem]{Example} 

\usepackage{multicol}
\usepackage{natbib}
\usepackage{pgf-pie}
\usepackage{pgfplots}
\usepackage{algorithm}
\usepackage{tikz}
\usepackage{pgfplots}
\usetikzlibrary{shadows}
\usepackage[edges]{forest}
\usetikzlibrary{shapes,arrows,positioning,calc,chains}
\pgfplotsset{
standard/.style={%Axis format configuration
axis x line=middle,
axis y line=middle,
enlarge x limits=0.15,
enlarge y limits=0.15,
every axis x label/.style={at={(current axis.right of origin)},anchor=north west},
every axis y label/.style={at={(current axis.above origin)},anchor=north east},
every axis plot post/.style={mark options={fill=white}}
}
}
\usepackage{booktabs}
\usepackage{verbatim}
\usepackage{subcaption}
\usepackage{graphicx} % support the \includegraphics command and options
\usepackage{multirow}
\usepackage{lipsum}

\makeatletter
\def\footnoterule{\relax%
\kern-5pt
\hbox to \columnwidth{\hfill\vrule width .9\columnwidth height 0.4pt\hfill}
\kern4.6pt}
\makeatother

\usepackage[bb=dsserif]{mathalpha}
\usepackage{bm}

\usepackage{color,hyperref}
\definecolor{darkblue}{rgb}{0.0,0.0,0.6}
\hypersetup{
colorlinks=true,       % false: boxed links; true: colored links
linkcolor=darkblue,    % color of internal links (change box color with linkbordercolor)
citecolor=darkblue,    % color of links to bibliography
filecolor=darkblue,    % color of file links
urlcolor=darkblue      % color of external links
}

\begin{document}

\begin{frontmatter}
%\runtitle{Insert a suggested running title}  % Running title for regular papers but only if the title is over 5 words. The running title is not shown in the output.

\title{From Semi-Infinite Constraints to Structured Robust Policies: Optimal Gain Selection for Financial Systems} % Title, preferably not more than 10 words.

\thanks[footnoteinfo]{This manuscript was supported in part by the National Science and Technology Council (NSTC), Taiwan, under Grant: NSTC113--2628--E--007--015--. It is a refined version of our earlier preprint~\cite{hsieh2022robust}, available on arXiv.}

\author[CHHSIEH]{Chung-Han Hsieh}\ead{ch.hsieh@mx.nthu.edu.tw.} 
%~$\;$ and $\;$    % Add the 
%\author[CHHSIEH]{B. Ross Barmish}\ead{barmish@engr.wisc.edu}               % e-mail address 
%\author[Baiae]{Publius Maro Vergilius}\ead{vergilius@culture.ir}  % (ead) as shown

\address[CHHSIEH]{Department of Quantitative Finance, National Tsing Hua University, Hsinchu, Taiwan, 300044.}
%\address[Barmish]{Senate House, Rome}             % full addresses
%\address[Baiae]{The White House, Baiae}        % here.

% keyword list or with the help of the Automatica keyword wizard

\begin{abstract}     
This paper studies the robust optimal gain selection problem for financial trading systems, formulated within a \emph{double linear policy} framework, which allocates capital across long and short positions.
The key objective is to guarantee \emph{robust positive expected} (RPE) profits uniformly across a range of uncertain market conditions while ensuring risk control. This problem leads to a robust optimization formulation with \emph{semi-infinite} constraints, where the uncertainty is modeled by a bounded set of possible return parameters. We address this by transforming semi-infinite constraints into structured policies---the \emph{balanced} policy and the \emph{complementary} policy---which enable explicit characterization of the optimal solution. Additionally, we propose a novel graphical approach to efficiently solve the robust gain selection problem, drastically reducing computational complexity. Empirical validation on historical stock price data demonstrates superior performance in terms of risk-adjusted returns and downside risk compared to conventional strategies. This framework generalizes classical mean-variance optimization by incorporating robustness considerations, offering a systematic and efficient solution for robust trading under uncertainty.
\end{abstract}

\end{frontmatter}

\section{Introduction} \label{SECTION: INTRODUCTION}
A financial trading system is deemed {\em robust} if it generates consistent returns regardless of market direction. This ability to consistently yield positive expected trading gains, irrespective of market direction, is a key metric for assessing a system's robustness and resilience to market volatility. This property is particularly critical given the frequent regime changes in financial markets and the inherent unpredictability of returns. Early foundational works related to robustness issues in financial systems can be found in \cite{dokuchaev2002dynamic} and \cite{barmish2008trading}.

In the robust control literature, the goal is often to ensure stability or performance guarantees across all values within a specified uncertainty set $\mathcal{U}$. In this paper, we study a discrete-time stochastic financial system where the input stochastic process has uncertain parameters mean $\mu$ and variance $\sigma^2.$  
Unlike typical stability or $\mathcal{H}_\infty$-norm constraints, we impose a novel \emph{robust positive expectation} (RPE) constraint, which guarantees financial viability by maintaining a nonnegative expected state across all admissible values of the uncertain parameters.

Several studies have applied {\em control-theoretic} approach to trading problems. For instance, ~\cite{zhang2001stock} presented an optimal selling rule, while~\cite{lee2007multiagent} examined multi-agent approaches to $q$-learning for stock trading. 
Other notable contributions include optimal pair trading strategies, see~\cite{tie2018optimal, deshpande2018generalization}, state positivity for delayed financial systems~\cite{hsieh2019positive},  and adaptive data-driven stock trading and prediction models, see \cite{chiang2016adaptive, abbracciavento2022data}. Additionally, \cite{van2015distributionally} studied distributionally robust control of constrained stochastic systems, while  \cite{barmish2024tutorial} presented a comprehensive tutorial on the control-theoretic approach to stock trading.

More recently,  stock trading using \emph{limit order book}\footnote{The limit order book is a record of all outstanding {\em limit orders} in a market, organized by price levels.} information were studied in \cite{xue2021optimization,barmish2021feedforward, nielsen2023characterizing}.
Among existing methodologies, a notable approach is the Simultaneous Long-Short (SLS) control scheme, see \cite{barmish2015new}, which guarantees {\em robust positive expectation} (RPE) across a wide array of stock price processes.
The SLS framework has inspired numerous extensions, including adaptations for time-varying stock price drifts and volatilities \cite{primbs2017robustness}, Merton's diffusion model~\cite{baumann2017simultaneously}, Proportional–Integral (PI) controllers \cite{malekpour2018generalization}, and discrete-time systems with delays \cite{malekpour2016stock}, among others.  In~\cite{deshpande2020simultaneous}, an SLS control with cross-coupling is proposed to trade two stocks. A generalized RPE theorem was presented in~\cite{o2020generalized}, which can have different parameters for the long and short sides of the trade, is studied. 
In~\cite{maroni2019robust}, a problem formulation that treats prices as disturbances was proposed, obtaining the SLS controller parameters as the solution of an~$\mathcal{H}_\infty$-optimization problem. However, the approach may result in an overly conservative feedback gain, highlighting the need for more flexible frameworks.

Despite significant progress made in control-theoretic trading strategies, a notable gap remains in systematically selecting optimal decision variables to maximize cumulative returns while ensuring robustness under uncertain market dynamics. Existing robust optimization methods~\cite{ben2009robust, blanchet2022distributionally} primarily focus on static problems with uncertainty modeled over fixed ambiguity sets, whereas semi-infinite optimization offers an  alternative for handling continuous uncertainty~\cite{stein2012solve, goberna2018recent}. However, its computational complexity poses significant challenges in financial applications.

Building on our earlier work~\cite{hsieh2022robust, hsieh2022robustness} on the double linear policy framework, this paper addresses the general robust gain selection problem under semi-infinite robustness constraints.  
By reformulating the semi-infinite constraints into two structured policies---the balanced policy and the complementary policy---we derive explicit characterizations of the optimal solution, establishing new results on existence and uniqueness.
Additionally, we propose a novel graphical approach that drastically reduces computational complexity compared to directly solving semi-infinite optimization problems. Unlike prior work, which primarily focused on static problems or specific scenarios, our approach generalizes classical mean-variance optimization, e.g., see \cite{markowitz1952portfolio}, by incorporating robustness considerations and providing explicit performance evaluation criteria, such as robust positive growth and monotonicity.

The primary contributions of this paper are as follows. First, we introduce a novel semi-infinite constrained {\em robust optimal gain selection problem} within a double linear policy framework. Our approach incorporates robust positive expected profits (RPE) as an additional robustness constraint, setting our work apart from~\cite{marandi2022extending}, which primarily considers parametric uncertainty sets.  Second, we show that semi-infinite RPE constraints can be reformulated into  two structured policies: the balanced policy and complementary policy, leading to new theoretical results on the existence and uniqueness of the optimal solution. 
Our analysis leverages optimization results, including Berge’s Maximum and Minimum Theorems, see \cite{aliprantis2006infinite} and the Extended Extreme Value Theorem, see \cite{beck2017first}, to rigorously establish the existence and uniqueness of optimal solutions under semi-infinite constraints.
Third, to overcome the computational burden of semi-infinite optimization,  we propose a novel graphical approach that efficiently solves the problem by exploiting key structural properties, significantly reducing~complexity. 

The rest of the paper is organized as follows.
In Section~\ref{SECTION: Problem Formulation}, we provide the problem formulation and necessary preliminaries, including a formal definition of the RPE property and the robust optimal gain selection problem. Section~\ref{SECTION: Robust Optimal Gain Selection} presents the main theoretical results for solving the problem under semi-infinite RPE constraints and characterizes the solution explicitly.  In Section~\ref{SECTION: Solving Robust Optimal Gain Selection}, we establish the existence and uniqueness of the optimal solution, followed by Section~\ref{SECTION: A Graphical Approach to Finding the Optimal Gain}, which describe a graphical method and algorithm for determining the optimal feedback gain.
Section~\ref{SECTION: Illustrative Examples} validates the proposed framework through numerical experiments using historical stock price data and Monte Carlo simulations. To handle dynamic market conditions, we apply a rolling horizon heuristic, inspired by \cite{blanchet2022distributionally, wang2022data}. Comparisons with a standard linear feedback strategy calibrated using a non-robust model reveal that our approach yields comparable or superior average returns while significantly reducing volatility.
Finally, Section~\ref{SECTION: Conclusion} concludes the paper with remarks on potential future research directions.

\section{Problem Formulation and Preliminaries}	\label{SECTION: Problem Formulation}
Let $(\Omega, \mathcal{F}, \mathbb{P})$ be a complete probability space.  Consider a sequence of independent random variables $X:=\{X(k)\}_{k\geq 0}$ representing {\em per-period returns}.\footnote{One can view these variables as independent samples drawn from a population with unknown distribution.} 
These variables are bounded almost surely within known limits, i.e.,
$ X(k) \in [X_{\min}, X_{\max}]$ for all $k$, with 
$
-1 < X_{\min} < 0 < X_{\max} < \infty,
$
and~$X_{\min}$ and~$X_{\max}$ are within the support of $X(k)$. 
Instead of assuming a single known mean~$\mathbb{E}[X(k)] := \mu  $  and  variance~${\rm var}(X(k)) := \sigma^2 \geq 0$, we allow $\mu$ and~$\sigma^2$ to be \emph{uncertain} and bounded within a known set~$\mathcal{U} \subset [X_{\min}, X_{\max}] \times \mathbb{R}_+$, where
\[
\mathcal{U} :=\{ (\mu, \sigma^2): \mu \in [\underline{\mu}, \ \overline{\mu}], \; \sigma^2 \in [0, \overline{\sigma}^2] \},
\]
and $-1< X_{\min} \leq \underline{\mu} <  0 < \overline{\mu} \leq X_{\max}$, with $\overline{\sigma}^2$ denoting an upper bound on the variance. In practice, $\mathcal{U}$ might be chosen as a box around an empirical estimate of $\mu$ and~$\sigma^2$, or it could represent a confidence region, such as a 95\% confidence interval.

\begin{remark} 
While our framework assumes independent returns for mathematical tractability, extending to non-iid returns would introduce significant complexity due to temporal dependencies. Such dependencies fundamentally alter the problem structure, rendering explicit closed-form or structured solutions generally infeasible. Nevertheless,  independence can serve as a practical ``first approximation'' trading model, see \cite{fama1965behavior}. Moreover, empirical results in Section~\ref{SECTION: Illustrative Examples} show that the proposed method remains effective when  applied to financial data with mild dependence; see also Section~\ref{SECTION: Conclusion} for future directions.
\end{remark}

\subsection{The Double Linear Policy and System Dynamics}
\begin{figure}[h!]
\centering
\includegraphics[width=0.95 \linewidth]{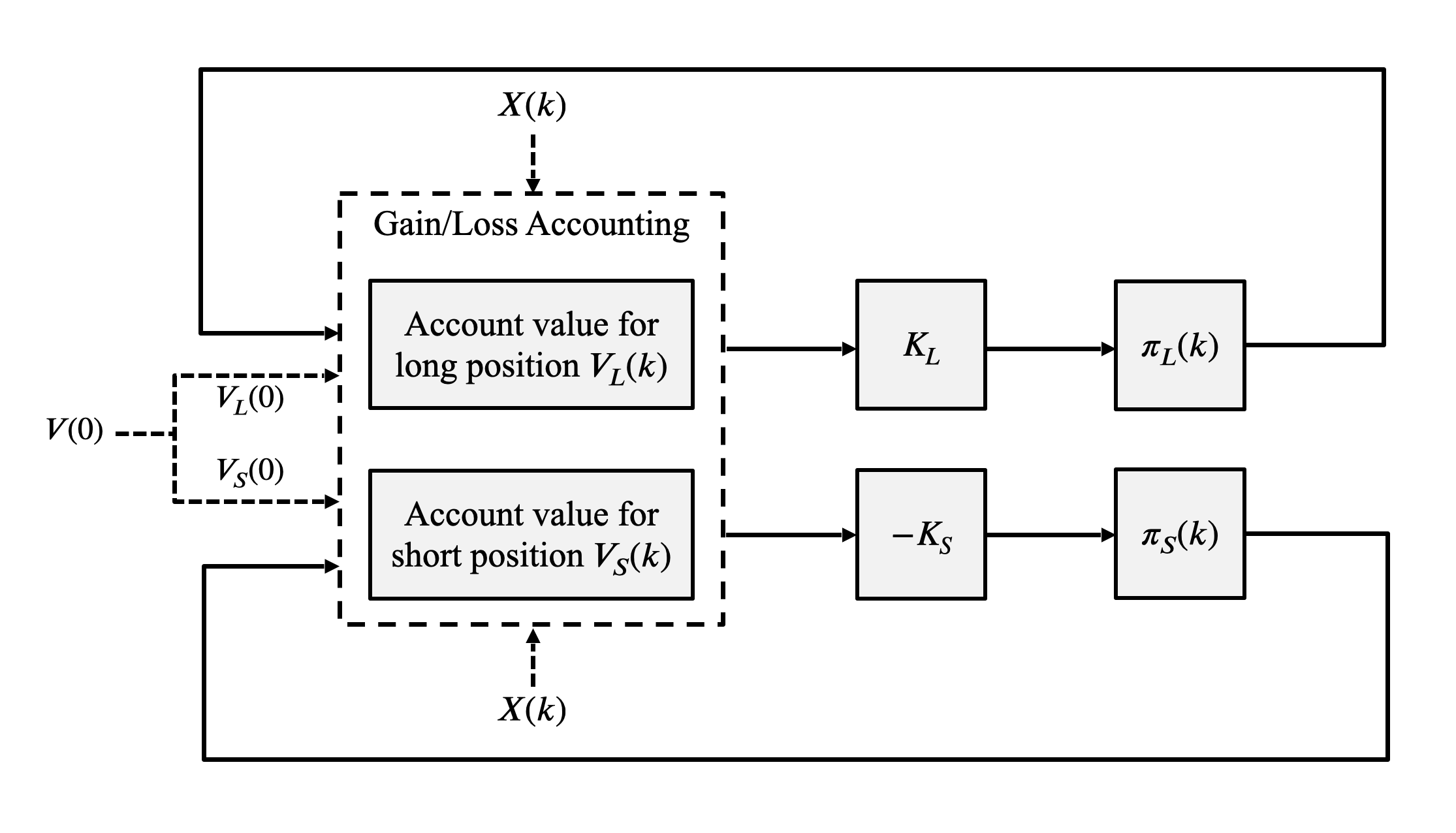}
\caption{Block Diagram of the Double Linear Policy Scheme.}
\label{fig:blockdiagram}
\end{figure}

Following~\cite{hsieh2022robust, wang2023robustness}, we consider a financial trading system starting with an initial state $V(0) := V_0 > 0$. 
The system has two components defined using a parameter $\alpha \in [0,1]$:  
$V_L(0) := \alpha V_0$ for the {\em long} position and $V_S(0) := (1 - \alpha)V_0$ for the {\em short} position.
At each stage, the {\em double linear policy}~$\pi(k)$ is expressed as the sum of two linear feedback components,~$\pi_L(k)$ and~$\pi_S(k)$. Specifically,~$
\pi(k) := \pi_L(k) + \pi_S(k),
$
where
$
\pi_L( k ) := K_L V_L( k )$ and
$
\pi_S( k ) :=  -  K_S V_S ( k ).
$
The admissible gains~$K_i$ for $i \in \{L, S\}$ belong to the interval~$\mathcal{K}:= \left[0, K_{\max}\right]$, with $K_{\max} := \min \left\{1,\, {1}/{X_{\max}} \right\} $.
Thus,  the parameter triple~$(\alpha, K_L, K_S) \in [0,1] \times \mathcal{K}\times \mathcal{K}$ fully characterizes the policy~$\pi(k)$. See Figure~\ref{fig:blockdiagram} for an illustration of the double linear policy~scheme.

\begin{remark} \rm
$(i)$ The double linear policy considered here generalizes the one used in~\cite{hsieh2022robustness} by allowing the long and short positions to have different weights.
$(ii)$ From a control-theoretic perspective, the double linear policy can be viewed as a discrete-time state-feedback controller with two channels---one for the ``long'' component and one for the ``short'' component---applied to the stochastic input $X(k)$. 
\end{remark}

The system {\em state}, or {\em account values}, $V(k)$ evolves according to the following stochastic discrete-time dynamics:
\[
\begin{cases}
{V_L}( k + 1 ) = {V_L}( k ) + X( k ){\pi_L}( k ) + (V_L(k) - \pi_L(k)) r_f(k);\\
{V_S}( k + 1 ) = {V_S}( k ) + X( k ){\pi_S}( k ).
\end{cases} 
\]
where $r_f(k) \geq 0$ for all $k$ is a {\em risk-free} rate for a bank account or a treasury bond. 
Therefore, the system state at stage $k$ is given~by
\begin{align} 
V(k) 
&= V_L(k) + V_S(k)  \notag \\
&= V_0\left(  \alpha R_+(k)  + (1-\alpha)R_-(k)  \right) \label{eq: account value eq}
\end{align} 
where
$
R_+(k) := \prod_{j = 0}^{k - 1} \left( 1 + K_L X( j ) + (1 - K_L) r_f(k) \right)$
and
$	
R_-(k) := \prod_{j = 0}^{k - 1} \left( 1 - K_S X( j ) \right).
$
Note that when~$r_f >0$, the state of $V_L(k)$ increases deterministically, leading to a surplus in profit and positively skewing performance. 
Henceforth, we assume without loss of generality that $r_f:=0$ to ensure fairness when studying roubst performance, as $r_f>0$ would introduce additional gains into the account value.
As mentioned in the previous subsection, every $K_i \in \mathcal{K}$ for $i \in \{L, S\}$  assures state positivity and cash-financed positions. That is, for all~$k$, the~$K$-value that can potentially lead to $V(k) < 0$ is disallowed.

%\medskip
\begin{lemma}[State Positivity and Cash-Finance] \label{lemma: state positivity}
For stage $k=0,1,\dots$, the double linear policy~$\pi(k)$ leads to all-time state positivity with probability one; i.e.,~$
\mathbb{P}(V(k) >0) = 1.
$
Moreover, the double linear policy is cash-financed, meaning that~$
| \pi(k) | \leq V(k)
$
for all~$K_i \in \mathcal{K}$ with $i \in \{L, S\}$ and $k \geq 1$ with probability~one.
\end{lemma}

\begin{proof}
See Appendix~\ref{Appendix: proofs in Problem Formulation}.
\end{proof}

\subsection{Cumulative Gain and Robust Positive Expectation}
Let $\mathcal{G}_k(  \alpha, K_L, K_S;  X ) := V(k) - V_0$ be the {\em cumulative gain-loss function} up to stage $k$.
Having defined the function, we now state the robust positive expectation (RPE) with risk control property formally.  

%\medskip
\begin{definition}[RPE with Risk Control] \rm \label{definition: RPE}
A double linear policy guarantees the  {\em robust positive expectation}~(RPE) with risk control property if, for each $(\mu, \sigma^2) \in \mathcal{U}$, there exists a triple~$(\alpha, K_L, K_S) \in [0, 1] \times \mathcal{K} \times \mathcal{K}$ such that,  for~$k >1$, the expected cumulative gain function satisfies
$$
\mathbb{E}\left[ \mathcal{G}_k(\alpha, K_L, K_S;  X) \right] \geq 0,
$$
and the standard deviation remains bounded by
$$
{\rm std}(\mathcal{G}_k(\alpha, K_L, K_S; X)) \leq s
$$
where $s$ is a constant representing a predefined risk threshold.
\end{definition}

%\medskip
\begin{remark} \rm
The term ``robust" indicates that the expected gain function remains nonnegative uniformly for all $\mu \in [\underline{\mu}, \overline{\mu}]$. This ensures that even if the actual mean~$\mu$ deviates---whether slightly negative or positive---the policy ensures a nonnegative expected gain.  This RPE property will later be incorporated as an expectation constraint in the robust optimal gain selection problem; see next subsection to follow. 
\end{remark}

\subsection{The Robust Optimal Gain Selection Problem}
The robust optimal gain selection problem requires that the \emph{worst-case expected gain} remain positive, and the \emph{worst-case standard deviation} stays below a threshold $s$.

%\medskip
\begin{problem}[Robust Optimal Gain Selection] \label{problem: robust optimal gain selection}
Fix~$N > 1$.	For stages $k= 2, 3, \dots, N$, given a constant~$s \in (0, s_{\max})$, where $s_{\max}$ is an upper bound on the standard deviation of the cumulative gain function,
%$$
%s_{\max}:= \sup_{(\mu, \sigma^2) \in \mathcal{U}} {\rm std}\left( \mathcal{G}_k(\tfrac{1}{2}, K_{\max}, K_{\max}; X) \right),
%$$ 
we seek  a triple  $(\alpha^*, K_L^*, K_S^*) \in [0,1] \times \mathcal{K} \times \mathcal{K}$ that solves the following robust maximization problem subject to semi-infinite RPE with risk control constraints: 
\begin{align}
& \max_{\alpha, K_L, K_S} \ \inf_{(\mu, \sigma^2) \in \mathcal{U}} \mathbb{E}[ \mathcal{G}_k (\alpha, K_L, K_S; X)] \\
&\; {\rm s. t.} \; \sup_{(\mu, \sigma^2) \in \mathcal{U}} {\rm std}\left( \mathcal{G}_k(\alpha, K_L, K_S; X) \right)  \leq s; \notag\\
&\qquad \inf_{(\mu, \sigma^2) \in \mathcal{U}} \mathbb{E}[ \mathcal{G}_k (\alpha, K_L, K_S; X)]  \geq 0.  \label{constraint: RPE constraint}
\end{align}
\end{problem}

\begin{remark}
$(i)$ Note that $\sigma^2$ does not influence $\mathbb{E}[{ \mathcal{G}}_k(\cdot)  ] $, although it does affect the variance. Hence, as shown in Section~\ref{SECTION: Robust Optimal Gain Selection}, the semi-infinite RPE constraints~\eqref{constraint: RPE constraint} reduce to analyzing only two structured policies with specific parametric triples, significantly simplifying the problem.
$(ii)$ Problem~\ref{problem: robust optimal gain selection} generalizes the classical mean-variance (MV) problem in three key ways: It incorporates dynamic stages, incorporates a double linear policy instead of a single feedback strategy, and enforces an RPE constraint for enhanced robustness. 
In the special case where $\underline{\mu} = \mu = \overline{\mu}$ and $\overline{\sigma}^2 = \sigma^2$, i.e., when there is no model uncertainty, it recovers the single-model problem described in \cite{hsieh2022robust}.
\end{remark}

\section{Robust Positivity and Structural Results} \label{SECTION: Robust Optimal Gain Selection}
This section presents the robustness analysis of the proposed framework, which is essential for solving the Robust Optimal Gain Selection Problem~\ref{problem: robust optimal gain selection}. We first derive closed-form expressions for the expected gain and variance  for any fixed $(\mu, \sigma^2)$.  These expressions serve as the foundation for analyzing the robust positive expectation (RPE) property and deriving structured policies.

\subsection{Expected Gain and Variance}

For any fixed pair $(\mu, \sigma^2)$, the following lemma provides closed-form expressions for the expected cumulative gain and its variance under the double linear policy.

\begin{lemma}[Nominal Expected Gain and Variance] \label{lemma: expected cumulative gain or loss and variance}
For stage $k=0,1,\dots$, the double linear policy, parameterized by the triple $(\alpha, K_L, K_S) \in [0,1] \times \mathcal{K} \times \mathcal{K}$, yields
the expected cumulative gain function
\begin{align*}
&\mathbb{E}[{\mathcal{G}}_k(\alpha, K_L, K_S; X)] \\
&= {V_0} \left(  \alpha \left( 1 + K_L \mu \right)^k + (1-\alpha)\left( 1 - K_S\mu \right)^k- 1 \right).
\end{align*}
Moreover, the corresponding variance is given by
\begin{align*}
& {\rm var}(\mathcal{G}_k (\alpha, K_L, K_S; X))\\
&= {\alpha^2 V_0^2}((1 + K_L\mu)^2 + K_L^2 \sigma^2)^k  \\
&+{(1-\alpha)^2 V_0^2} ( (1 - K_S \mu)^2 + K_S^2\sigma^2)^k\\
&\quad +{ 2\alpha (1-\alpha)V_0^2 }  (1 + \mu (K_L- K_S) -K_L K_S (\sigma^2+ \mu^2))^k\\
&\quad -2\alpha V_0^2 (1+K_L\mu)^k-2(1-\alpha)V_0^2 (1-K_S\mu)^k+V_0^2\\
&\quad -{V_0}^2 \left( \alpha \left( 1 + K_L \mu  \right)^k+ (1-\alpha)\left( 1 - K_S \mu  \right)^k - 1 \right)^2.
\end{align*}
Of course, the standard deviation of the cumulative gain function is 
$$
{\rm std}(\mathcal{G}_k(\alpha, K_L, K_S; X)) =\sqrt{{\rm var}(\mathcal{G}_k(\alpha, K_L, K_S; X))}.
$$
\end{lemma}

\begin{proof}
See Appendix~\ref{Appendix: proofs in Main Results}.
\end{proof}

\begin{remark} \rm
$(i)$ As a sanity test, if no trading occurs; i.e., $K_L = K_S=0$,  then
$
\mathbb{E}[{\mathcal{G}}_k(\alpha, K_L, K_S; X)] = {\rm var}(\mathcal{G}_k(\alpha, K_L, K_S; X))=0.
$
$(ii)$~If the returns have no uncertainty; i.e., $\sigma^2 =0$,  a straightforward calculation leads to
$
{\rm var}(\mathcal{G}_k(\alpha, K_L, K_S; X)) =0.
$
$(iii)$ If the returns have no trend; i.e., $\mu = 0$,  it follows that~$\mathbb{E}[{\mathcal{G}}_k(\alpha, K_L, K_S; X)] =0$, while the variance simplifies~to
\begin{align*}
& {\rm var}(\mathcal{G}_k (\alpha, K_L, K_S; X))\\
&\qquad = {\alpha^2 V_0^2}( K_L^2 \sigma^2)^k  +{(1-\alpha)^2 V_0^2} ( K_S^2\sigma^2)^k\\
&\quad \qquad +{ 2\alpha (1-\alpha)V_0^2 }  (1 -K_L K_S \sigma^2)^k - V_0^2.
\end{align*}
\end{remark}

\subsection{Characterization of Robust Positive Expectation} \label{SECTION: Robust Positive Expectation Property and Its Ramifications}
We now characterize the robust positive expectation (RPE) property of the double linear policy under the uncertainty set~$\mathcal{U}$.  Since $\sigma^2$ only affects the variance but not the expected value, the RPE constraint reduces to ensure that $\mathbb{E}[{ \mathcal{G}}_k(\alpha, K_L, K_S; X) ] \geq 0$ for all  $\mu \in [\underline{\mu}, \overline{\mu}]$.

\begin{theorem}[Characterization of RPE] \label{theorem: Robust Positive Expectation}
Consider the double linear policy with triple $(\alpha, K_L, K_S)  \in (0,1) \times \mathcal{K} \times \mathcal{K}$.  For all $k \geq 1$,  the robust positive expectation (RPE) constraint
\begin{align}
\inf_{(\mu, \sigma^2) \in \mathcal{U}} \mathbb{E} [ {\mathcal{G}}_k(\alpha, K_L, K_S; X) ] \geq 0  \label{ineq: robust postiive expectation property}
\end{align}
holds if and only if the double linear policy takes one of the following two structured forms:
\begin{itemize}
\item[$(i)$] 	\textit{Balanced Policy}:  $\alpha = \tfrac{1}{2}$ and $K_L = K_S = K \in \mathcal{K}$.
\item[$(ii)$] 	\textit{Complementary Policy}: $\alpha \in (0,1)$ and $K_L = 1-\alpha$ and $K_S = \alpha$.
\end{itemize}
\end{theorem}

\begin{proof} 
According to Lemma~\ref{lemma: expected cumulative gain or loss and variance}, $\mathbb{E}[\mathcal{G}_k(\cdot)]$ only depends on $\mu$. Specifically, it follows that	
\[
\inf_{(\mu, \sigma^2) \in \mathcal{U}} \mathbb{E} [ {\mathcal{G}}_k(\alpha, K_L, K_S; X) ]  = \min_{\mu \in [\underline{\mu}, \overline{\mu}]} \mathbb{E} [ {\mathcal{G}}_k(\alpha, K_L, K_S; X) ]
\]
where the minimum is attained since $\mathbb{E} [ {\mathcal{G}}_k(\alpha, K_L, K_S; X) ]$ is a continuous function over compact interval $[\underline{\mu}, \overline{\mu}].$
Having established this, we now prove sufficiency. In particular,  the balanced policy with triple $(\alpha, K_L, K_S) = (\tfrac{1}{2}, K, K)$ with $K \in \mathcal{K}$  implies RPE constraint~\eqref{ineq: robust postiive expectation property}, as shown in Lemma~\ref{lemma: PRE for Balanced Policy}. On the other hand, the complementary policy with triple $(\alpha, K_L, K_S) = (\alpha, 1-\alpha, \alpha)$ with $\alpha \in (0,1)$ implies RPE constraint~\eqref{ineq: robust postiive expectation property}, as shown in Lemma~\ref{lemma: RPE for Complementary Case}.  

For necessity, we proceed with a proof by contradiction: Assuming that $
\mathbb{E}[  {\mathcal{G}}_k(\alpha, K_L, K_S; X)] \geq 0  $
for all $\mu \in [\underline{\mu}, \overline{\mu}]$ and all integer $k \geq 1$ and assume neither the balanced policy nor complementary policy hold. Specifically, the negation, via De Morgan's Law, leads to the following three cases.

\textit{Case 1}. For $\alpha =0,$ $K_L, K_S \notin (0,1)$ with $K_L \neq K_S$. In this case, 
\begin{align}  
\mathbb{E}[  {\mathcal{G}}_k(\alpha, K_L, K_S; X)]
&=  {V_0}   (    \left( 1 - K_S\mu \right)^k- 1  ),
\end{align}
then there exists $\mu \in (0, \overline{\mu})$ such that $	\mathbb{E}[  {\mathcal{G}}_k(\alpha, K_L, K_S; X)] <0$, which is contradicting to~\eqref{ineq: robust postiive expectation property}.

\textit{Case 2}.  On the other hand, for $\alpha = 1$, $K_L, K_S \notin (0,1)$ with $K_L \neq K_S$. In this case, 
\begin{align}  
\mathbb{E}[  {\mathcal{G}}_k(\alpha, K_L, K_S; X)]
&=  {V_0}   (    \left( 1 + K_L\mu \right)^k- 1  ).
\end{align}
Then there exists $\mu \in (\underline{\mu}, 0)$ such that $	\mathbb{E}[  {\mathcal{G}}_k(\alpha, K_L, K_S; X)] <0$, which is contradicting to~\eqref{ineq: robust postiive expectation property}.

\textit{Case 3.} For $\alpha \in (0,1)$ with either $K_L \neq 1-\alpha$ or $K_S \neq \alpha$, 
we note that the expected cumulative gain function is given by
$
\mathbb{E}[{\mathcal{G}}_k(\alpha, K_L, K_S; X)] 
= {V_0}(\alpha (1 + K_L\mu)^k + (1-\alpha)(1 - K_S\mu)^k - 1).
$
At $\mu=0$, the above expression evaluates to zero.  The derivative with respect to $\mu$ is
\[
\frac{\partial}{\partial \mu} \mathbb{E}[{\mathcal{G}}_k(\alpha, K_L, K_S; X)]
\big|_{\mu=0}
= V_0 \, k  (\alpha K_L - (1-\alpha)K_S ).
\]
For the robust positive expectation condition~\eqref{ineq: robust postiive expectation property} to hold \emph{near} $\mu=0$, the function cannot dip below zero on either side of $\mu=0$. Consequently, its derivative at~$\mu=0$ must vanish, i.e.\ $\alpha K_L = (1-\alpha)K_S$. Otherwise, if $\alpha K_L \neq (1-\alpha)K_S$, the derivative can be positive or negative. Suppose $\alpha K_L > (1-\alpha)K_S$, then for small negative $\mu$, the term $(1 + K_L\,\mu)^k$ will shrink faster (since $K_L$ is relatively large), while $(1 - K_S\,\mu)^k$ will not compensate sufficiently, driving the sum below 1, hence making $\mathbb{E}[{\mathcal{G}}_k(\alpha,K_L,K_S;X)]<0$. Conversely, if $\alpha K_L < (1-\alpha)K_S$, then for small positive $\mu$ the same argument applies: $(1 - K_S\,\mu)^k$ shrinks faster than $(1 + K_L\,\mu)^k$ can offset, again dropping below 1. In either scenario, we obtain a negative value for the expected gain, contradicting \eqref{ineq: robust postiive expectation property}.

Hence, to avoid violating RPE near $\mu=0$, we must have $\alpha K_L = (1-\alpha)K_S$, i.e.\ $\frac{K_L}{1-\alpha} = \frac{K_S}{\alpha}$. 
If in addition $K_L \neq 1-\alpha$ or $K_S \neq \alpha$, then
$
\frac{K_L}{1-\alpha} 
= 
\frac{K_S}{\alpha} 
\neq 1.
$
One can then choose a small positive or negative $\mu$ (depending on whether $K_L > 1-\alpha$ or $K_L < 1-\alpha$) so that $(1 + K_L \mu)^k + (1 - K_S \mu)^k < 1$. This forces the expectation to be negative and thus contradicts the RPE~property. 
\end{proof}

\begin{remark}
Theorem~\ref{theorem: Robust Positive Expectation} shows that the robust positive expectation (RPE) constraint simplifies to two structured policies: the balanced policy and the complementary policy. These two policies represent two distinct trading strategies: $(i)$ a balanced approach, which allocates capital symmetrically between long and short positions initially, and $(ii)$ a complementary approach, which adjusts the allocation asymmetrically based on market conditions. This result is significant because it drastically reduces the complexity of the robust optimal gain selection problem by narrowing the search space to two low-dimensional structured policies, making the problem tractable while retaining robustness across the entire uncertainty set $(\mu, \sigma^2) \in \mathcal{U}$. 
\end{remark}

\begin{lemma}[RPE for Balanced Policy] \label{lemma: PRE for Balanced Policy} 
Consider the double linear policy with $(\alpha, K_L, K_S)  = (1/2, K, K)$ and~$K \in \mathcal{K}$.  For all $k > 1$,  the expected cumulative gain satisfies
$
\mathbb{E}[ {\mathcal{G}}_k(\frac{1}{2}, K, K; X)] > 0
$
for all~$K \in \mathcal{K}$ with~$K > 0$ and all $\mu\in[\underline{\mu},\overline{\mu}]\setminus\{0\}$. Moreover,  if $K \mu =0$, then $\mathbb{E}[ {\mathcal{G}}_k(\frac{1}{2}, K, K; X)] = 0.$
\end{lemma}

%\begin{proof}
%	See Appendix~\ref{Appendix: Technical Proofs}.
%\end{proof}

\begin{proof}%[Proof of Lemma~\ref{theorem: Robust Positive Expectation}] 
The proof is elementary.	Fix $k>1$ and $K \in \mathcal{K}$. With~$(\alpha, K_L, K_S)  = ( \tfrac{1}{2}, K, K)  $, Lemma~\ref{lemma: expected cumulative gain or loss and variance} tells us that
\[
\mathbb{E}[ {\mathcal{G}}_k (\tfrac{1}{2}, K, K; X)] = \tfrac{V_0}{2} ( { (  1 + K\mu    )^k} + {\left(  1 - K\mu   \right)^k} - 2 ).
\]
If $K \mu = 0$, which occurs when either $K=0$ or $\mu =0$, it is straightforward to see that $\mathbb{E}[ {\mathcal{G}}_k(\frac{1}{2}, K, K; X)]=0.$
On the other hand, if~$\mu \neq 0$ and $K>0$, then $K\mu \neq 0$. 
By the basic fact that 
$
(1+x)^k + (1-x)^k > 2
$
for all $x \neq 0$ and~$k > 1$, the desired positivity is guaranteed.
\end{proof}

\begin{remark} \rm 
$(i)$ Lemma~\ref{lemma: PRE for Balanced Policy} indicates that by fixing a single~$\alpha^*=\frac{1}{2}$, the RPE property is preserved for all admissible~$K>0$.  
This fact implies that one may have an extra degree of freedom to select an ``optimal"~$K$ without affecting the desired positivity. Hence, a ``separation design" is seen; see \cite{chen2012linear}.
$(ii)$ If~$\alpha \neq \frac{1}{2}$ and~$K_L = K_S=K$,  the RPE property may fail to hold; see the next example.
\end{remark}

\begin{example}[Uneven Alpha Leads to a Failing RPE] \rm
Fix stage $k=2$ and initial state $V_0=1$. Consider a double linear policy with~$(\alpha, K, K) = \tfrac{1}{4}  \times \mathcal{K}  \times \mathcal{K}$ for $K>0$.  
According to Lemma~\ref{lemma: expected cumulative gain or loss and variance},  the corresponding expected cumulative gain is given by
\begin{align*}
\mathbb{E}\left[ {\mathcal{G}}_2 (\tfrac{1}{4}, K, K; X  ) \right ]
&=   \tfrac{1}{4} \left( 1 + K\mu \right)^2 + \tfrac{3}{4} \left( 1 - K\mu \right)^2- 1\\
& =  - K\mu (1 -K\mu ).
\end{align*}
If $\mu \in (0,  X_{\max})$, it follows that $0 < K\mu < 1$. 
Hence, we have $	\mathbb{E}[{\mathcal{G}}_2(\frac{1}{4}, K, K; X)] <0.$ 
For example, if~$K\mu = 1/4$, then~$\mathbb{E}[ {\mathcal{G}}_2(\frac{1}{4}, K, K; X)] \approx -0.188 <0 $; i.e., an expected  trading {\em loss} is seen. 
Additionally, by fixing an admissible~$K>0$, we can see that the RPE fails up to some~$k$. For example, take the same parameter $\alpha = 1/4$, if $K =  \mu = \frac{1}{2}$, then~$\mathbb{E}[ { \mathcal{G}}_k (\tfrac{1}{4},\tfrac{1}{2}, \tfrac{1}{2}; X)] < 0$ for $1 \leq k \leq 5$; see Figure~\ref{fig: illustrative ex for failing robust G}.

\pgfplotsset{width=6.5cm,compat=1.8}
\begin{figure}[htbp]
\centering
\begin{tikzpicture}
	\begin{axis}[%
		standard,
		domain = 1: 7,
		samples = 7,
		xlabel={$k$},
		ylabel={$\mathbb{E}[ { \mathcal{G}}_k (\tfrac{1}{4}, \tfrac{1}{2}, \tfrac{1}{2}; X )]$}]
		\addplot+[ycomb, blue, variable =\k, line width=1pt, opacity = 0.3] {1/4*(1+1/4)^\k + 3/4*(1-1/4)^\k-1};
	\end{axis}
\end{tikzpicture}
\caption{Uneven $\alpha$ May Lead to Failing RPE. A Visualization of $\mathbb{E}[ { \mathcal{G}}_k  ( \tfrac{1}{4}, \tfrac{1}{2}, \tfrac{1}{2}; X  )]$. }
\label{fig: illustrative ex for failing robust G}
\end{figure} 

\end{example}

\begin{lemma}[RPE for Complementary Policy] \label{lemma: RPE for Complementary Case}
Consider the double linear policy with triple $(\alpha, K_L, K_S)  = (\alpha, 1-\alpha, \alpha)$ for $\alpha \in (0,1)$.  For all $k \geq 1$,  the expected cumulative gain satisfies
$
\mathbb{E}[{\mathcal{G}}_k(\alpha, 1-\alpha, \alpha; X)]  \geq 0
$
for all~$\mu \in [\underline{\mu}, \overline{\mu}]$.
\end{lemma}

\begin{proof}
Let $\alpha \in (0,1)$, $K_L = 1- \alpha$ and $K_S = \alpha$.
We note that 
\begin{align*}
&\mathbb{E}[ {\mathcal{G}}_k(\alpha, 1-\alpha, \alpha; X)] \\
&=  {V_0}   (  \alpha  \left( 1 +(1-\alpha) \mu \right)^k +  (1-\alpha) \left( 1 - \alpha \mu \right)^k- 1  ).
\end{align*}
Since $V_0 >0$, it suffices to prove $f(\mu):= \alpha  \left( 1 +(1-\alpha) \mu \right)^k +  (1-\alpha) \left( 1 - \alpha \mu \right)^k- 1 \geq 0$ for all $\mu \in [\underline{\mu}, \overline{\mu}] $ and integer $k\geq 1$.
Compute the derivative 
 \begin{align*}
f'(\mu)  
%&= \alpha k (1-\alpha) (1 + (1-\alpha)\mu)^{k-1}  -\alpha (1-\alpha) k (1 - \alpha \mu)^{k-1} \\
&= \alpha k (1-\alpha) [(1 + (1-\alpha)\mu)^{k-1}  -  (1 - \alpha \mu)^{k-1}  ].
\end{align*}
Note that $\alpha k (1-\alpha) \geq 0$ because $\alpha \in (0,1)$. Hence,  the sign of $f'(\mu)$ depends on the expression inside the brackets.

\textit{Case 1.} $\mu \geq 0$.
In this case, $(1 + (1 -\alpha) \mu) \geq (1-\alpha \mu)$ since $(1-\alpha) \mu \geq 0. $ Therefore, $(1 + (1-\alpha)\mu)^{k-1}  \geq  (1 - \alpha \mu)^{k-1}$ and we have $f'(\mu) \geq 0.$

\textit{Case 2.} $\mu < 0$.
In this case, $(1 + (1 -\alpha) \mu) < (1-\alpha \mu)$ since $(1 -\alpha) \mu <0.$ Therefore, $(1 + (1-\alpha)\mu)^{k-1}  <  (1 - \alpha \mu)^{k-1}$ and we have $f'(\mu) \leq 0$. 

In combination of the two cases above, we conclude that~$f'(\mu) \geq 0$ for $\mu \geq 0$ and $f'(\mu)  \leq 0$ for $\mu < 0$. Therefore, $f(\mu)$ attains its minimum at $\mu = 0.$ It then remains to evaluate the $f(\mu)$ at $\mu = 0$.  Indeed,
\begin{align*}
f(0)  = \alpha  \cdot 1^k +  (1-\alpha) \cdot 1^k- 1 = 0.
\end{align*}
Therefore, $f(\mu)$ attains minimum at $\mu = 0$ with $\min_\mu f(\mu) = 0$, which implies that $f(\mu) \geq 0$ for all $\mu$ and all $k \geq 1.$
\end{proof}

\subsection{Monotonicity Lemmas}
In this subsection, we focus on a monotonicity property of the double linear policy, which is critical for determining the optimality in the Robust Optimal Gain Selection Problem~\ref{problem: robust optimal gain selection}. The first result states that any \emph{balanced} double linear policy with~$\alpha = \tfrac{1}{2}$ ensures a property stronger than robust positive expectation: the expected cumulative gain function is both positive and monotonically increasing in~time~$k$. The second result indicates that the monotonicity of both the expected value and variances also holds with respect to the feedback gain $K$.

\begin{lemma}[Robust Growth of Balanced Policy] \label{corollary: Robust Expected Growth}
Consider the balanced double linear policy with triple $(\alpha, K, K) \in \{ \tfrac{1}{2}\} \times \mathcal{K} \times \mathcal{K}$.
Then, for~$k \geq 1$, the expected cumulative gain function satisfies
$$
\mathbb{E}\left[ {\mathcal{G}}_{k+1}\left( \tfrac{1}{2}, K, K; X \right) \right] \geq 
\mathbb{E}\left[ {\mathcal{G}}_k\left( \tfrac{1}{2}, K, K; X \right) \right]
$$ 
for  all $K \in \mathcal{K}$ and all $(\mu, \sigma^2) \in \mathcal{U}$.
\end{lemma}

\begin{proof}
See Appendix~\ref{Appendix: proofs in Main Results}.
\end{proof}

\begin{lemma}[Monotonicity of Balanced Policy]  \label{lemma: robust monotonicity for iid case}
Fix an integer $k>1$.  
Consider the balanced double linear policy with $(\alpha, K, K) \in \{\tfrac12\} \times \mathcal{K} \times \mathcal{K}$.  
For each $(\mu,\sigma^2) \in \mathcal{U}$, we have
\begin{enumerate}
\item The expected cumulative gain
$
\mathbb{E} [\mathcal{G}_k(\tfrac{1}{2}, K,K;\,X) ]
$
is nondecreasing in $K\in\mathcal{K}$.  
Moreover, if $\mu\neq0$, it is \emph{strictly} increasing for $k>1$.
\item The variance
$
\mathrm{var} \left( \mathcal{G}_k(\tfrac12,K,K;\,X) \right)
$
is nondecreasing in $K\in\mathcal{K}$.  
Moreover, if $\sigma^2>0$, it is \emph{strictly} increasing for $k>1$.
\end{enumerate}
\end{lemma}

\begin{proof}
See Appendix~\ref{Appendix: Technical Proofs}.
\end{proof}

\begin{corollary} 
\label{cor: worst case monotonicity}
Adopt the same setting as Lemma~\ref{lemma: robust monotonicity for iid case}.  Then:
\begin{align*}
&	K
\;\mapsto\;
\inf_{(\mu,\sigma^2)\,\in\,\mathcal{U}}
\mathbb{E}\bigl[\mathcal{G}_k(\tfrac12,K,K;\,X)\bigr]
\quad\text{and}\quad \\
&	K
\;\mapsto\;
\sup_{(\mu,\sigma^2)\,\in\,\mathcal{U}}
\mathrm{var}\bigl(\mathcal{G}_k(\tfrac12,K,K;\,X)\bigr)
\end{align*}
are also nondecreasing (and strictly increasing if some~$\mu\neq 0$, $\sigma^2>0$). 
% In particular, the worst-case standard deviation
%\[
%K \mapsto \sup_{(\mu,\sigma^2)\,\in\,\mathcal{U}}
%{\rm std}\bigl(\mathcal{G}_k(\tfrac12,K,K;\,X)\bigr)
%\]
%is nondecreasing in $K$, and strictly increasing if there exists at least one $(\mu,\sigma^2)\in\mathcal{U}$ with $\sigma^2>0$ and $k>1$.
\end{corollary}

\begin{proof} 
Adopt the same setting as Lemma~\ref{lemma: robust monotonicity for iid case}. 
The nondecreasingness of 
$\inf_{(\mu,\sigma^2) \in \mathcal{U}}
\mathbb{E}[ \mathcal{G}_k( \tfrac12,K, K; X) ]$ and $\sup_{(\mu,\sigma^2)\,\in\,\mathcal{U}}
\mathrm{var}(\mathcal{G}_k( \tfrac12,K,K;\,X))$ in $K$ follows by combining monotonicity with respect to $K$ (pointwise in $\mu$ and $\sigma^2$) with the infimum or supremum. Specifically, if $f(K) \leq g(K)$ pointwise for all $(\mu, \sigma^2) \in \mathcal{U}$, then $\inf f(K) \leq \inf g(K)$ and $\sup f(K) \leq \sup g(K)$.  
\end{proof}

%\begin{remark} \label{theorem: strictly increasing and convexity of variance}  
%	Consider a double linear policy with triple $(\alpha, K_L, K_S) \in (0,1) \times \mathcal{K} \times \mathcal{K}$ with $K>0.$
%The variance 
%		  $\mathrm{var} (\mathcal{G}_k(\alpha,K_L,K_S; X) )$ is \emph{strictly increasing} in~$\sigma^2 \in [0, \overline{\sigma}^2]$ and is \emph{convex} in $\mu \in [\underline{\mu}, \overline{\mu}]$.
%This implies that the maximum is attained at boundary points of the uncertainty set. That is,  
%	\begin{align*}
%	&\sup_{(\mu, \sigma^2) \in \mathcal{U}} {\rm std} ( \mathcal{G}_k(\alpha, K_L, K_S; X)  ) \\
%	&= \max \begin{Bmatrix}
%		\ {\rm std} ( \mathcal{G}_k(\alpha, K_L, K_S; X)|_{\mu = \underline{\mu}, \sigma^2 
%			=\overline{\sigma}^2},   
%		\\
%		{\rm std} (  \mathcal{G}_k(\alpha, K_L, K_S; X)|_{\mu = \overline{\mu}, \sigma^2 =\overline{\sigma}^2}  
%	\end{Bmatrix} .
%\end{align*} 
%\end{remark}

\section{Solving Robust Optimal Gain Selection } \label{SECTION: Solving Robust Optimal Gain Selection}  
According to Theorem~\ref{theorem: Robust Positive Expectation},  the semi-infinite RPE constraint~\eqref{constraint: RPE constraint} holds if and only if the double linear policy  with parameters $(\alpha, K_L, K_S)\in [0,1] \times \mathcal{K} \times \mathcal{K}$ satisfies either one of the two structured forms:

\begin{itemize}
\item \textit{Balanced Policy:} $\alpha = \tfrac{1}{2}$ and $K_L = K_S = K \in \mathcal{K}$. 
\item \textit{Complementary Policy:} $\alpha \in (0,1)$ and $K_L = 1-\alpha$, $K_S = \alpha$.
\end{itemize}

Thus, we define the feasible sets for the two structured policies as follows: 
\[
\Theta_B: = \left\{ (\alpha, K_L, K_S) : \alpha = \tfrac{1}{2}, K_L = K_S, K \in \mathcal{K} \right\}
\]
and
\[
\Theta_C = \left\{ (\alpha, K_L, K_S)  : K_L = 1-\alpha, K_S = \alpha, \alpha \in [0,1] \right\}
\]
Then define the overall feasible set $\Theta:= \Theta_B \cup \Theta_C$. Since the variance $\sigma^2$ only affects the standard deviation but not the sign of the mean, the RPE constraint~\eqref{definition: RPE} holds if and only if the triple $(\alpha, K_L, K_S) \in \Theta$.  Consequently, the robust optimal gain selection~problem can be reformulated as the following equivalent problem:

%\medskip
\begin{problem}[Equivalent Robust Optimal Gain Selection] \label{problem: robust optimal gain selection_generalized}
Fix~$N > 1$.	For stage $k= 2, 3, \dots, N$, given a constant $s \in (0, s_{\max})$, where 
$$
s_{\max}:={\rm std}(\mathcal{G}_k(\tfrac{1}{2}, K_{\max}, K_{\max}; X) ),
$$ 
we seek  a triple $(\alpha^*, K_L^*, K_S^*) \in [0,1] \times \mathcal{K} \times \mathcal{K}$ that solves the following constrained maximization problem: 
\begin{align*}
& \max_{\alpha, K_L, K_S} \; \inf_{(\mu, \sigma^2) \in \mathcal{U}} \mathbb{E}[ {\mathcal{G}}_k (\alpha, K_L, K_S; X)] \\
&\; {\rm s. t.} \; (\alpha, K_L, K_S ) \in \Theta\\
& \qquad  \sup_{(\mu, \sigma^2) \in \mathcal{U}}  {\rm std}({\mathcal{G}}_k (\alpha, K_L, K_S; X)) \leq s.
\end{align*}
\end{problem}

\begin{remark} \rm
Unlike classical robust control approaches, such as those based on linear matrix inequalities (LMI), this formulation allows us to exploit the monotonicity of the standard deviation with respect to the feedback gain $K$ to find an explicit solution. This leads to a significant computational reduction relative to generic semi-infinite optimization techniques.
\end{remark}

\subsection{Existence and Uniqueness} \label{subsection: Existence and Uniqueness}
The following theorem establishes the existence and uniqueness of the optimal solution for the balanced case.

\begin{theorem}[Existence and Uniqueness of Optimal Balanced Policy]\label{theorem: Existence and Uniqueness of Optimal Solution} %[Existence and Uniqueness] 
For  {$1<k \leq N$}, consider the double linear linear policy with triple $(\alpha, K_L, K_S) \in \Theta_B$. Given the targeted standard deviation $s \in (0, s_{\max})$ with 
\begin{align} \label{eq: s_max}
s_{\max} = \sup_{(\mu, \sigma^2) \in \mathcal{U}} {\rm std}( {\mathcal{G}}_k (\tfrac{1}{2}, K_{\max}, K_{\max}; X)).
\end{align}
Then there exists a solution $(\alpha, K_L, K_S) = (\frac{1}{2}, K^*, K^*)$  that solves the Robust Optimal Gain Selection Problem~\ref{problem: robust optimal gain selection_generalized}. Moreover, if $\mu \neq 0$ and $\sigma >0$, then the solution is unique.
\end{theorem}

\begin{proof}  
Since $ \mu \neq 0$ and $k>1$,  Theorem~\ref{theorem: Robust Positive Expectation} implies that $\alpha = \frac{1}{2}$ guarantees the desired RPE property for all admissible $K \in \mathcal{K}$ with $K > 0$ and $\mu \in [\underline{\mu}, \overline{\mu}]$. 

Hence, to complete the proof, it remains to determine~$K^*$.
Fix constant $s \in (0, s_{\max})$ and  integer $1< k \leq N$. 
We begin by noting that the standard deviation $ {\rm std}({\mathcal{G}}_k(\frac{1}{2}, K; K; X))$ is  continuous in $K \in \mathcal{K}$ with $\mathcal{K}$ being compact; hence, by Berge's Maximum Theorem, $f(K):=\sup_{(\mu, \sigma^2) \in \mathcal{U}} {\rm std}({\mathcal{G}}_k(\frac{1}{2}, K; K; X))$ is lower semicontinuous. Hence, the feasible set $\mathcal{F}_B:=\{K \in \mathcal{K}: f(K) \leq s\}$  is a closed and bounded set, which is compact.

Now note that the objective function $h_B(K) := \inf_{(\mu, \sigma^2) \in \mathcal{U}} \mathbb{E}[ \mathcal{G}_k (\tfrac{1}{2}, K, K; X)]$, which is an infimum over continuous functions over compact uncertainty set $\mathcal{U}$. By Berge's minimum theorem, see \cite{aliprantis2006infinite}, $h_B(K)$ is upper semicontinuous, which is defined on the compact feasible set $\mathcal{F}_B \cap \Theta_B$. By the Extended Extreme Value Theorem, see \cite{beck2017first},~$h_B(K)$ attains its maximum at some~$ K^* \in \mathcal{F}_B \cap \Theta_B$. Moreover,
$$
0 = {\rm std}\left(  {\mathcal{G}}_k(\tfrac{1}{2}, 0, 0; X) \right) < s <   s_{\max}.
$$
%where $s_{\max} = \sup_{(\mu, \sigma^2) \in \mathcal{U}} {\rm std}( {\mathcal{G}}_k(\frac{1}{2}, K_{\max}, K_{\max}; X).$
Next, we show the uniqueness of $K^*$. Since  $\mu \neq 0$ and~$\sigma >0$, Lemma~\ref{lemma: robust monotonicity for iid case} tells us that $\sup_{(\mu, \sigma^2) \in \mathcal{U}} {\rm var}(\mathcal{G}_k(\cdot))$  is strictly increasing in $K$; hence, so is the standard deviation, implying that~$K^*$ is unique. 
Thus,  combining the results established above, we conclude that the triple~$(\alpha, K_L, K_S) = (\frac{1}{2}, K^*, K^*)$ is the unique solution to the robust optimal gain selection problem under the balanced policy structure.  
\end{proof}

\begin{remark} \rm
The $ s_{\max} $ in Equation~\eqref{eq: s_max}  is well-defined and finite. To see this, note that ${\rm std}(\mathcal{G}_k(\cdot) )$, as shown later in Section~\ref{SECTION: Robust Optimal Gain Selection}, is continuous function of~$\mu$ and $\sigma^2$ over a compact set~$\mathcal{U}$. Hence, by the Weierstrass Extreme Value Theorem, the maximum~$s_{\max}$  is attained.
The following theorem addresses the existence of an optimal solution for the complementary policy.
\end{remark}

\begin{theorem}[Existence of Optimal Complementary Policy] \label{theorem: Existence of Optimal Complementary Policy}
For $(\alpha, K_L, K_S) \in  \Theta_C$ and $k > 1$ and $\mu \neq 0$. Then Problem~\ref{problem: robust optimal gain selection_generalized} reduces to the following nonconvex program:
\begin{align} \label{problem: equivalent problem for complementary case}
& \max_{\alpha} \inf_{(\mu, \sigma^2) \in \mathcal{U}} \mathbb{E}[ {\mathcal{G}}_k (\alpha, 1-\alpha, \alpha; X)] \\
&\; {\rm s. t.}   \;  \sup_{(\mu, \sigma^2) \in \mathcal{U}} {\rm std}({\mathcal{G}}_k (\alpha, 1-\alpha, \alpha; X)) \leq s, \notag
\end{align}
for $s \in (0, s_{\max})$ for some $s_{\max}$,
which admits a solution~$\alpha^* \in [0, 1].$
\end{theorem}

\begin{proof}   
	The existence proof for the complementary policy follows a similar structure of the existence part of Theorem~\ref{theorem: Existence and Uniqueness of Optimal Solution}. However, for the sake of completeness, we provide a full proof here, noting that much of the reasoning mirrors Theorem~\ref{theorem: Existence and Uniqueness of Optimal Solution}, differing primarily in the parameterization by~$\alpha$ and the lack of uniqueness due to non-convexity. Specifically, we begin by defining the constraint function as
\[
f(\alpha) := \sup_{(\mu, \sigma) \in \mathcal{U}} \text{std}(\mathcal{G}_k(\alpha, 1 - \alpha, \alpha; \mu, \sigma)).
\]
By Lemma~\ref{lemma: expected cumulative gain or loss and variance}, it is readily seen that $\text{std}(\mathcal{G}_k (\cdot))$ is continuous in all its variables. Since  the uncertainty set $\mathcal{U}$ is compact, by {Berge's Maximum Theorem}, see \cite[Theorem 17.31]{aliprantis2006infinite}, $f(\alpha)$ is lower semicontinuous.  
Let the feasible set $\mathcal{F}_C$ be defined as $
\mathcal{F}_C = \{ \alpha \in [0, 1] : f(\alpha) \leq s \}.
$
Since $f(\alpha)$ is lower semicontinuous and~$[0, 1]$ is compact, by \cite[Theorem 2.6]{beck2017first}, $\mathcal{F}_C$ is a closed and bounded subset of $[0, 1]$, hence compact. Moreover, $\mathcal{F}_C$ is non-empty because $s \in (0, s_{\max})$ ensures that at least one feasible policy parameter $\alpha$ exists.

%\noindent \textbf{Step 2: Existence of an optimal solution.}  
Next, we study the objective function. Define
\[
h(\alpha) := \inf_{(\mu, \sigma) \in \mathcal{U}} \mathbb{E}[\mathcal{G}_k(\alpha, 1 - \alpha, \alpha; \mu, \sigma)].
\]
According to Lemma~\ref{lemma: expected cumulative gain or loss and variance}, the expected cumulative gain function $\mathbb{E}[ \mathcal{G}_k(\cdot)]$ is continuous in all variables and $\mathcal{U}$ is compact.  By {Berge's Minimum Theorem}, $h(\alpha)$ is upper semicontinuous. Moreover, since $h(\alpha)$ is defined on the compact feasible set~$\mathcal{F}_C \cap \Theta_C$, $h(\alpha)$ attains its maximum on~$\mathcal{F}_C\cap \Theta_C$ by the {Extended Extreme Value Theorem}, see \cite{beck2017first}.
Thus, there exists an optimal parameter~$\alpha^* \in \mathcal{F}_C \cap \Theta_C$
%$
%\alpha^* = \arg\max_{\alpha \in \mathcal{F}\cap \Theta_C} h(\alpha),
%$
which solves the robust gain selection problem under the complementary policy.
\end{proof}

\begin{remark} \rm
Since the objective function considered in Theorem~\ref{theorem: Existence of Optimal Complementary Policy} is not strictly concave, the uniqueness of the optimal complementary policy may not be guaranteed.  
\end{remark}

\section{Graphical Approach to Finding the Optimum} \label{SECTION: A Graphical Approach to Finding the Optimal Gain}  
Building on the theoretical results established in Section~\ref{SECTION: Robust Optimal Gain Selection}, this section introduces a graphical method to determine the optimal feedback gain efficiently. Semi-infinite constrained robust optimization problems typically require discretizing the uncertainty set or using iterative schemes, which can be computationally intensive. However, by leveraging the specific structure of our problem---particularly the monotonicity properties (see Lemma~\ref{lemma: robust monotonicity for iid case}) and the reduction to two structured policies (see Theorem~\ref{theorem: Robust Positive Expectation})---the proposed graphical approach focuses on evaluating low-dimensional curves in the mean-standard deviation plane. This avoids reliance on second-order cone programming (SOCP) or linear matrix inequalities (LMIs), improving computational feasibility for large horizons $N$.

\subsection{Parameterization for the Balanced Policy}
For  the balanced double linear policy framework, given~$\alpha = \frac{1}{2}$ and an integer~$1 < k \leq N$, defines a mapping
$$
K \mapsto \begin{pmatrix}
\sup_{(\mu, \sigma^2) \in \mathcal{U}} {\rm std}(\mathcal{G}_k(\tfrac{1}{2}, K, K; X)) \\
\inf_{(\mu, \sigma^2) \in \mathcal{U}} \mathbb{E}[ {\mathcal{G}}_k(\tfrac{1}{2}, K, K; X)]   
\end{pmatrix} \in \mathbb{R}^2.
$$
As the feedback gain~$K$ varies, the resulting point $({\rm std}(\mathcal{G}_k (\frac{1}{2}, K; X)),\; \mathbb{E}[ {\mathcal{G}}_k(\frac{1}{2}, K, K; X))]  \in \mathbb{R}^2$ traces out a {\em plane curve} $\mathcal{C}_B$, parameterized by $K$, in the mean-standard deviation plane. That is
\[
\mathcal{C}_B :=  \left\{  \begin{pmatrix}
\sup_{(\mu, \sigma^2) \in \mathcal{U}} {\rm std}(\mathcal{G}_k(\tfrac{1}{2}, K, K; X)) \\
\inf_{(\mu, \sigma^2) \in \mathcal{U}} \mathbb{E}[ {\mathcal{G}}_k(\tfrac{1}{2}, K, K; X)]   
\end{pmatrix} : K \in \mathcal{K} \right\}. 
\]
%where $\mathbb{R}_+^2 :=\{(x_1, x_2)\in\mathbb R^2: x_i \geq 0, i=1,2\}$.
Note that when~$K=0$, it implies 
$$
({\rm std}(\mathcal{G}_k(\alpha, 0, 0; X)), \mathbb{E}[ {\mathcal{G}}_k(\alpha, 0, 0; X))] = (0, 0),
$$
which corresponds to origin on $\mathbb{R}^2$.
If $\mu \neq 0$ and~$\sigma > 0$,  Monotonicity Lemma~\ref{lemma: robust monotonicity for iid case} implies that the plane curve~$\mathcal{C}$ generated by map 
$$
K \mapsto \left( {\rm std}(\mathcal{G}_k(\tfrac{1}{2}, K, K; X)), \mathbb{E}[ {\mathcal{G}}_k(\tfrac{1}{2}, K, K; X)] \right)
$$
is strictly increasing for all $K \in \mathcal{K}$.

\subsection{Parameterization for the Complementary Policy}
Recalling that a complementary policy is defined by $(\alpha, K_L, K_S) := (\alpha, 1-\alpha, \alpha)$ with $\alpha \in [0, 1]$. As in the balanced case, define for each integer $k \geq 1$ the mapping
\[
\alpha \mapsto \begin{pmatrix} 
&\sup_{(\mu, \sigma^2) \in \mathcal{U}} {\rm std}( \mathcal{G}_k ( \alpha, 1-\alpha, \alpha; X)) \\
&\inf_{(\mu, \sigma^2) \in \mathcal{U}} \mathbb{E}[  \mathcal{G}_k ( \alpha, 1-\alpha, \alpha; X)]  \\
\end{pmatrix} \in \mathbb{R}^2.
\]
Note that the worst-case standard deviation ${\rm std}( \mathcal{G}_k ( \alpha, 1-\alpha, \alpha; X)) $ is a continuous function of $\alpha, \mu,\sigma^2$. Taking the supremum over $\mu, \sigma^2 \in \mathcal{U}$ yields a continuous function $\alpha \mapsto \sup_{(\mu, \sigma^2) \in \mathcal{U}} (\mathcal{G}_k)$. On the other hand, the worst-case mean $\mathbb{E}[ \mathcal{G}_k(\alpha, 1-\alpha, \alpha; X)]$ depends on $\mu$, and $\sigma^2$ does not affect the mean. Therefore, $\inf_{(\mu, \sigma^2) \in \mathcal{U}} \mathbb{E}[\mathcal{G}_k(\cdot)] = \min_{ \mu \in [\underline{\mu}, \overline{\mu}]} \mathbb{E}[\mathcal{G}_k(\cdot)]$, which is also continuous in $\alpha.$ 
Thus, as $\alpha$ moves from $0$ to $1$, we trace out a second curve in the $(\mu, \sigma)$-plane. Denote this curve~by
{\small \[
\mathcal{C}_C = \left\{ \begin{pmatrix} 
&\sup_{(\mu, \sigma^2) \in \mathcal{U}} {\rm std}( \mathcal{G}_k ( \alpha, 1-\alpha, \alpha; X)) \\
&\inf_{(\mu, \sigma^2) \in \mathcal{U}} \mathbb{E}[  \mathcal{G}_k ( \alpha, 1-\alpha, \alpha; X)]  \\
\end{pmatrix} : \alpha \in [0, 1] \right\}
\]
}
\begin{remark} \rm
Unlike the balanced policy, the complementary policy mapping may not be strictly monotonic in $\alpha$. As a result, the curve for $\alpha \in [ 0,1]$  could be ``wiggly'' or bend back on itself. However, one can still compute these points numerically by discretizing $\alpha \in [0, 1]$ and applying standard optimization techniques. 
\end{remark}

\subsection{Merging Balanced and Complementary Policies}
Having obtained the balanced-policy curve $\mathcal{C}_B$, parameterized by $K$ and the complementary-policy curve $\mathcal{C}_C$, parameterized by $\alpha$, we merge the two in the mean-standard deviation plane. The result is two distinct ``frontiers'', and we define the ``efficient frontier'' as the \emph{upper envelope} of the two curves:
\[
\mathcal{C}_{\rm efficient} := \left\{ \begin{aligned}
& \text{All points in }\mathcal{C}_B \cup \mathcal{C}_C \text{ that are  \emph{not} }\\
& \text{dominated by other points}
\end{aligned} \right\}
\]
where a point is dominated if there exists another point with the same or lower standard deviation but a strictly higher mean, or vice versa.
In other words, we trace the union of the two curves and discard any parts that are strictly dominated. The remaining curve represents the efficient frontier, i.e., the set of all $({\rm std}, {\rm mean})$-pairs that cannot be improved in both directions.

Once the target standard deviation parameter $s \in (0, s_{\max})$ is specified, 
we can locate the maximum achievable value of the expected gain $\mathbb{E}[ {\mathcal{G}}_k(\cdot)]$ by intersecting the efficient frontier $\mathcal{C}_{\rm efficient}$ with the vertical line $\mathrm{std}({\mathcal{G}}_k(\cdot)) = s$.

\subsection{Graphical Approach and Algorithm} 
Once the target standard deviation parameter $s \in (0, s_{\max})$ is specified, 
we locate the maximum achievable value of the expected gain $\mathbb{E}[ {\mathcal{G}}_k(\cdot)]$ by intersecting the efficient frontier $\mathcal{C}_{\rm efficient}$ with the vertical line $\mathrm{std}({\mathcal{G}}_k(\cdot)) = s$.   
For the balanced policy curve $\mathcal{C}_B$, monotonicity allows for an explicit inverse computation. For the complementary policy curve $\mathcal{C}_C$, one proceeds by applying a binary-search or grid search method over $\alpha.$
The detailed procedures are summarized in Algorithm~\ref{algorithm: Gain Loss Optimization Algorithm}.
Throughout, $\mathbb{E}[\cdot]$ denotes the \emph{worst-case expected value} over all~$(\mu, \sigma^2)\in\mathcal{U}$, i.e.\ $\inf_{(\mu,\sigma^2)\in\mathcal{U}} \mathbb{E}[\cdot]$. Similarly, $\mathrm{std}(\cdot)$ denotes $\sup_{(\mu,\sigma^2)\in\mathcal{U}}\mathrm{std}( \cdot)$.

\begin{algorithm}[htbp] \small
\caption{Robust Gain Selection}
\label{algorithm: Gain Loss Optimization Algorithm}
\begin{algorithmic}[1]
\Require Initial state $V_0$, horizon $N$, target standard deviation $s 
\in (0, s_{\max})$, maximum return $X_{\max}$ and maximum gain $K_{\max}$, and uncertainty set $\mathcal{U} $.
\Ensure Optimal $(\alpha^*, K_L^*, K_S^*)$ and worst-case expected gain $\mathbb{E}[ {\mathcal{G}}_N(\alpha^*, K_L^*, K_S^*; X)]$

\State Initialize an empty list of feasible solutions: $\text{FeasibleSolutions} \leftarrow [\ ]$

\Statex \textbf{Step 1: Computing Balanced Policy}
\State Fix $\alpha = \frac{1}{2}$ for the balanced policy.
\State Solve:
\[
\begin{aligned}
\max_{K \in [0, K_{\max}]} & \quad \mathbb{E}[ {\mathcal{G}}_N\left(\tfrac{1}{2}, K, K; X \right)] \\
\text{s.t.} & \quad {\rm std}\left(\mathcal{G}_N\left(\tfrac{1}{2}, K, K; X\right)\right) \leq s
\end{aligned}
\]
\State Let the solution be \( K^* \) (obtained via grid search or a root-finding method).
\If{$K^*$ exists}
\State Compute $\mathbb{E}[ {\mathcal{G}}_N\left(\tfrac{1}{2}, K^*, K^*; X \right)]$
\State Add $\left(\tfrac{1}{2}, K^*, K^*\right)$ to $\text{FeasibleSolutions}$
\EndIf

\Statex \textbf{Step 2: Computing Complementary Policy}
\State Solve:
\[
\begin{aligned}
\max_{\alpha \in [0, 1]} & \quad \mathbb{E}[ {\mathcal{G}}_N\left(\alpha, 1 - \alpha, \alpha; X \right)] \\
\text{s.t.} & \quad {\rm std}\left(\mathcal{G}_N\left(\alpha, 1 - \alpha, \alpha; X\right)\right) \leq s
\end{aligned}
\]
\State Let the solution be \( \alpha^* \) (obtained via enumerating $\alpha$ or a line search).
\If{$\alpha^*$ exists}
\State Set \( K_L^* = 1 - \alpha^* \), \( K_S^* = \alpha^* \)
\State Compute $\mathbb{E}[ {\mathcal{G}}_N\left(\alpha^*, K_L^*, K_S^*; X \right)]$
\State Add $\left(\alpha^*, K_L^*, K_S^*\right)$ to $\text{FeasibleSolutions}$
\EndIf

\Statex \textbf{Step 3: Selecting Optimal Solution}
\If{$\text{FeasibleSolutions}$ is not empty}
\State From $\text{FeasibleSolutions}$, select the entry with the highest  $\mathbb{E}[ {\mathcal{G}}_N]$
\State Set $(\alpha^*, K_L^*, K_S^*)$ to the corresponding decision variables.
\State \Return Optimal $(\alpha^*, K_L^*, K_S^*)$ and $\mathbb{E}[ {\mathcal{G}}_N(\alpha^*, K_L^*, K_S^*; X)]$
\Else
\State \Return No feasible solution satisfies the standard deviation constraint.
\EndIf
\end{algorithmic}
\end{algorithm}

\section{Illustrative Examples}\label{SECTION: Illustrative Examples}\rm
To demonstrate the computational efficiency and robustness of our approach, we compare the proposed double linear policy with traditional single linear feedback policies. We first present a Monte-Carlo simulation example to illustrate the superior robustness of our approach, followed by an empirical study using historical data and rolling optimization to showcase adaptive robust performance in a real world setting.

\subsection{Monte-Carlo Simulation}
\begin{figure}[htbp] %These figure is generated using matlab code: Mean_Var_plot.m
\centering
\includegraphics[width=1\linewidth]{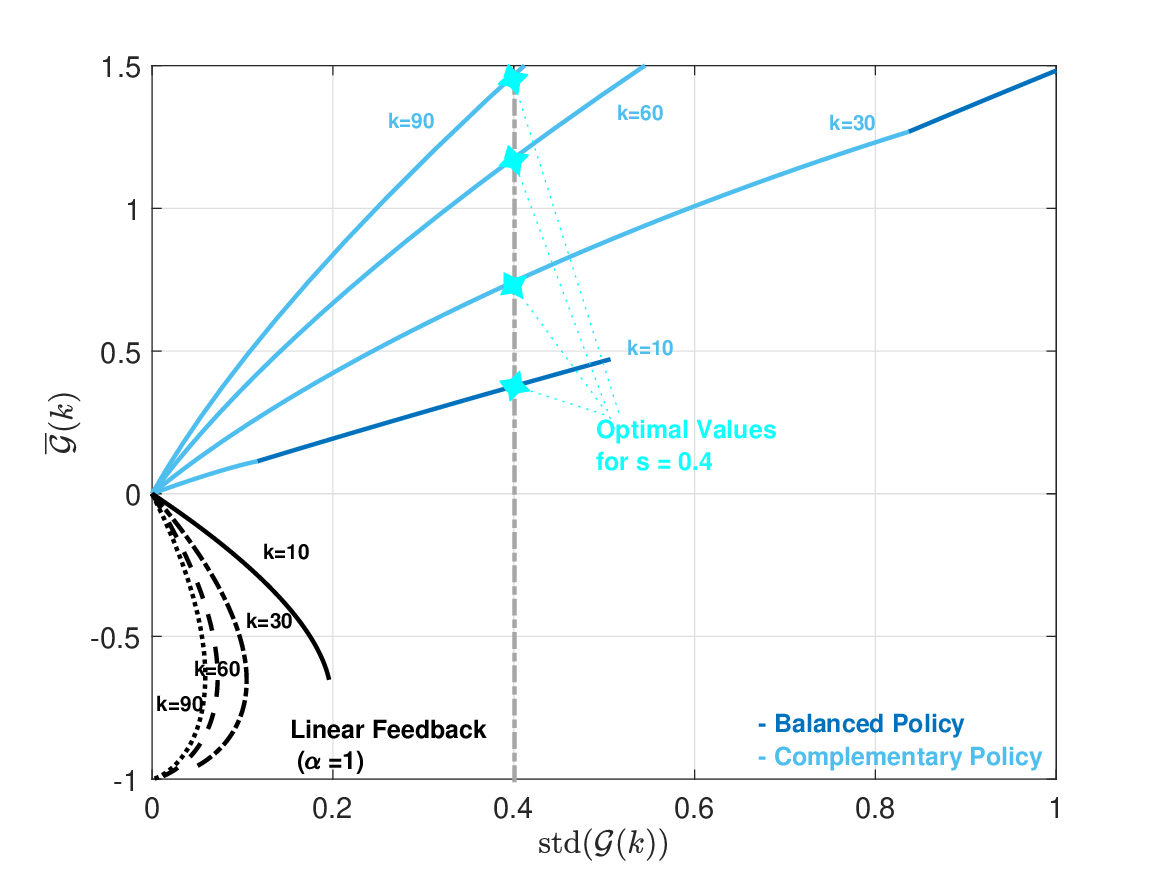}
\caption{Efficient Frontiers on Mean-Standard Deviation Plane: Double Linear Policy Versus Standard Linear Feedback. }
\label{fig:balancedvslinear}
\end{figure}

We begin with a simple illustrative example. Let the initial system state be~$V_0 := 1$.  We consider a hypothetical stock price process, where the independent returns~$X(k) \in (-1, 1]$  have a  mean~$\mu := -0.1$ and a standard deviation~$\sigma := 0.15$, representing a downward trending stock.\footnote{The parameters $\mu$ and $ \sigma $ are ``typical" estimates in annualized sense;  see~\cite{luenberger2013investment} for detailed discussion. }
By varying~$K \in \mathcal{K} = [0, 1]$, we plot the corresponding mean-standard deviation curves for different values of~$k \in \{ 10, 30, 60, 90\}$ in Figure~\ref{fig:balancedvslinear}. The blue curves represent the balanced policies and the cyan curves represent the complementary policy, while the single linear feedback policy (equivalent to the double linear policy with~$\alpha = 1$) is shown by the black curves.  From the figure, we see that the expected cumulative gain functions are positive for~$K>0$, consistent with the theoretical results. 

Suppose the target standard deviation is set to~$s := 0.4$,  corresponding to the gray vertical (dotted)  line ${\rm std}({\mathcal{G}}_k(\frac{1}{2}, K, K; X)):=s$ depicted in Figure~\ref{fig:balancedvslinear}. 
The optimal expected cumulative gain is determined by the intersection of the mean-standard deviation curves with the vertical line. As shown in the figure, the corresponding optimal expected cumulative gains are approximately $\mathbb{E}[ {\mathcal{G}}_k( \tfrac{1}{2}, K, K; X )] \approx 0.38, 0.73, 1.17, 1.38$ for~$k = 10, 30, 60, 90$, respectively.  
%The inverse exists since the expected cumulative gain function is strictly increasing in~$K \in \mathcal{K}$. 
The detailed optimal triples~$(\alpha^*, K_L^*, K_S^*)$ are summarized in Table~\ref{table: policy_parameters}.

Notably, in this example, the proposed double linear policy consistently outperforms the classical single linear feedback policy. Specifically, for any~$k>1$ and the same feasible standard deviation level $s > 0$,  the double linear policy yields a higher expected cumulative gain for all admissible gain values $K >0$.

\begin{table}[htbp] %These number below were computed using matlab code: Mean_Var_plot.m
\caption{Optimal Policy Parameters for Different Values of $k$.}
\label{table: policy_parameters}
\centering
\begin{tabular}{c c c c c }
\toprule
\textbf{$k$} & \textbf{$\alpha^*$} & \textbf{$K_L^*$} & \textbf{$K_S^*$} & {Optimal Policy Type} \\ \midrule
10           & 0.5              & 0.117         & 0.117         & Balanced Policy      \\ 
30           & 0.393           & 0.607         & 0.393         & Complementary Policy \\ 
60           & 0.359           & 0.642         & 0.359         & Complementary Policy \\ 
90           & 0.371           & 0.629         & 0.371         & Complementary Policy \\ 
\bottomrule
\end{tabular}

\end{table}

\begin{figure}[htbp]
\centering
\includegraphics[width=1\linewidth]{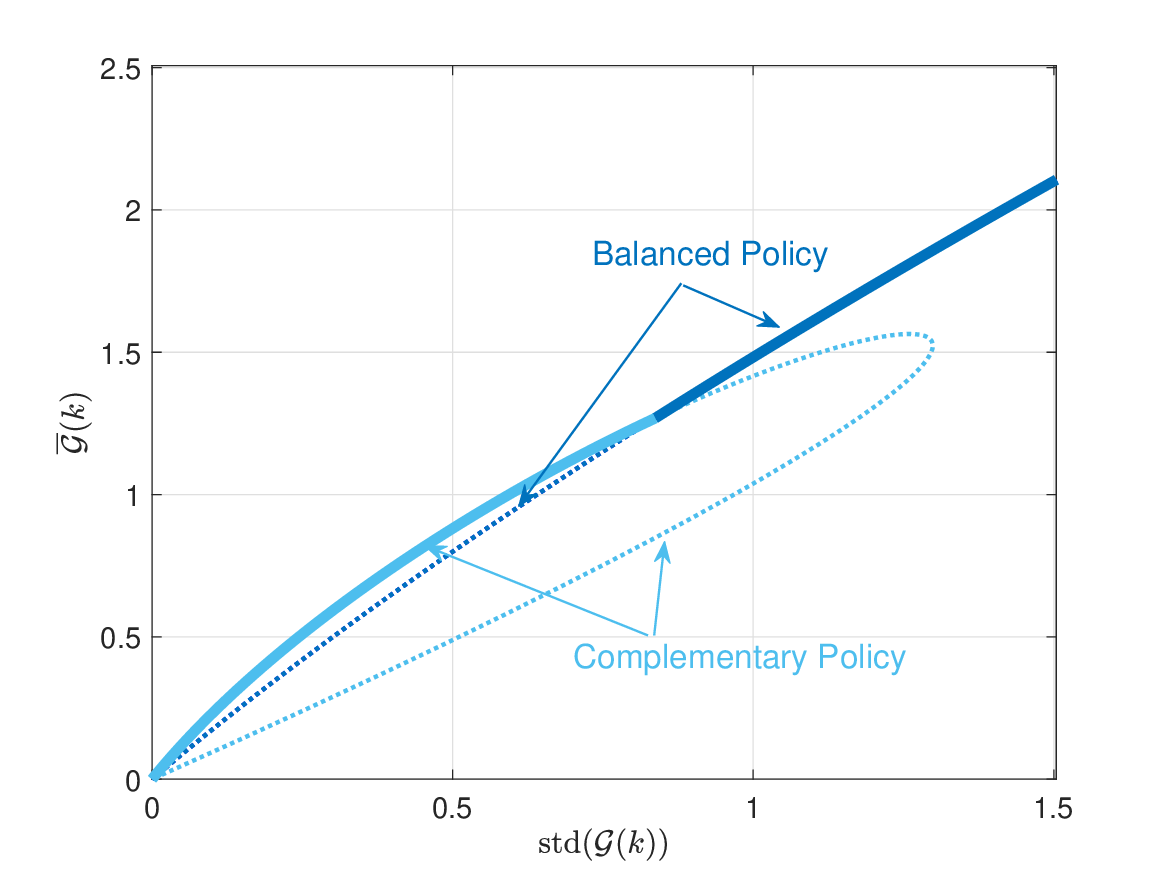}
\caption{Mean-Standard Deviation Curves: Balanced Policy and Complementary Policy. }
\label{fig:meanstdbalancedcomplementary}
\end{figure}

Figure~\ref{fig:meanstdbalancedcomplementary} shows one of the efficient frontiers from Figure~\ref{fig:balancedvslinear}, composed of the curves generated by the balanced policy and complementary policies. Note that  the dotted curve for the complementary policy shows non-convex behavior, particularly at higher variance levels. This confirms the theoretical findings about the complementary policy being nonconvex in $\alpha.$

Any upper-envelope intersection between the two curves suggests that the policies may provide comparable trade-offs in certain ranges of gain or variance.
If the solid line (balanced policy) dominates over the dotted line in a particular regime, it implies that the balanced policy may be more suitable for lower risk tolerance. Conversely, when risk tolerance exceeds a certain threshold, the complementary policy may provide better performance.
Beyond a specific variance level, however, the balanced policy may maintain its optimality due to its strict monotonicity properties.

\subsection{Empirical Studies: Rolling  Optimization} \label{SECTION: Trading With Historical Data}
In this subsection, we demonstrate the applicability of our theory using historical stock price data, 
specifically, for Tesla Motors~(ticker:~TSLA) over a 10-year period from January~02,~2015 to January~02,~2025, involving approximately $m=2,520$ trading days. 
Let~$s(k)$ denote the daily adjusted closing prices, and define the associated realized returns as
$
x(k) = ( s(k+1) - s(k) ) / s(k)
$
for stages~$k = 1, 2, \dots, m$ where $ m$ represents a rolling window size.

Suppose the tolerance for the standard deviation is set to ${\rm std}(\mathcal{G}_{k}(\frac{1}{2}, K, K; X)) \leq s := 0.1$. We perform rolling optimization with a window size of $m=60$ trading days to update the optimal triple $(\alpha, K_L, K_S)$. This approach mirrors adaptive robust control paradigms, where one periodically re-estimates uncertain parameters $\mu, \sigma^2$ and recalculates the policy's parameters to maintain the performance guarantees.

\begin{figure}[htbp] %This figure is generated using ``Double_Linear_Policy_TSLA.ipynb''
\centering
\includegraphics[width=1\linewidth]{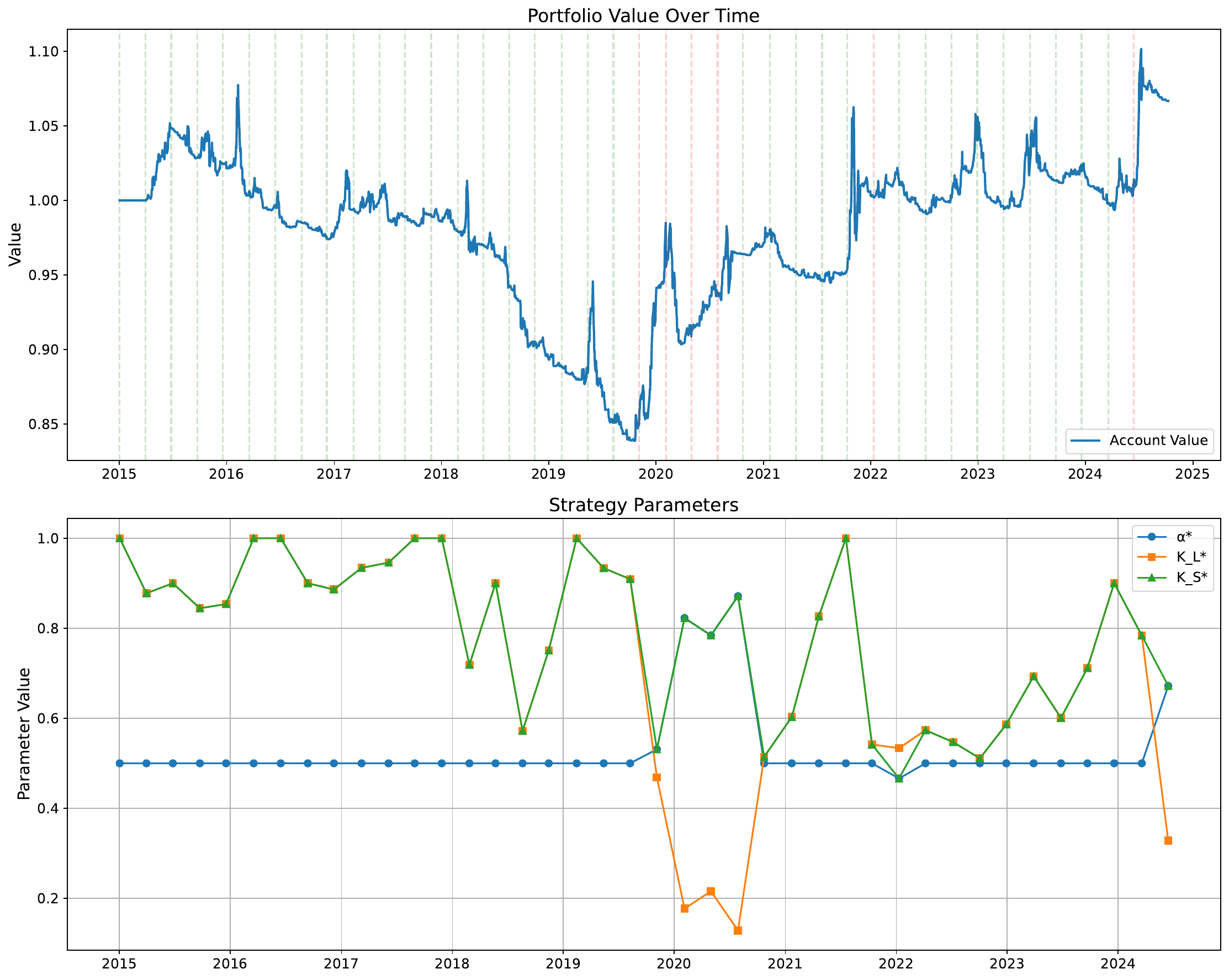}
\caption{Out-of-Sample Trading Performance with TSLA from~2015 to 2025 Using Rolling Optimization with a 60-Day Window Size: Upper Panel Shows Portfolio Value Over Time; Lower Panel Depicts Time-Varying Optimal Triples for the Double Linear Policy.}
\label{fig: tslatradingresults}
\end{figure}

\begin{figure}[htbp] %Code: on colab, runing  ``Copy of Double_Linear_Policy_TSLA.ipynb''
\centering
\includegraphics[width=0.91\linewidth]{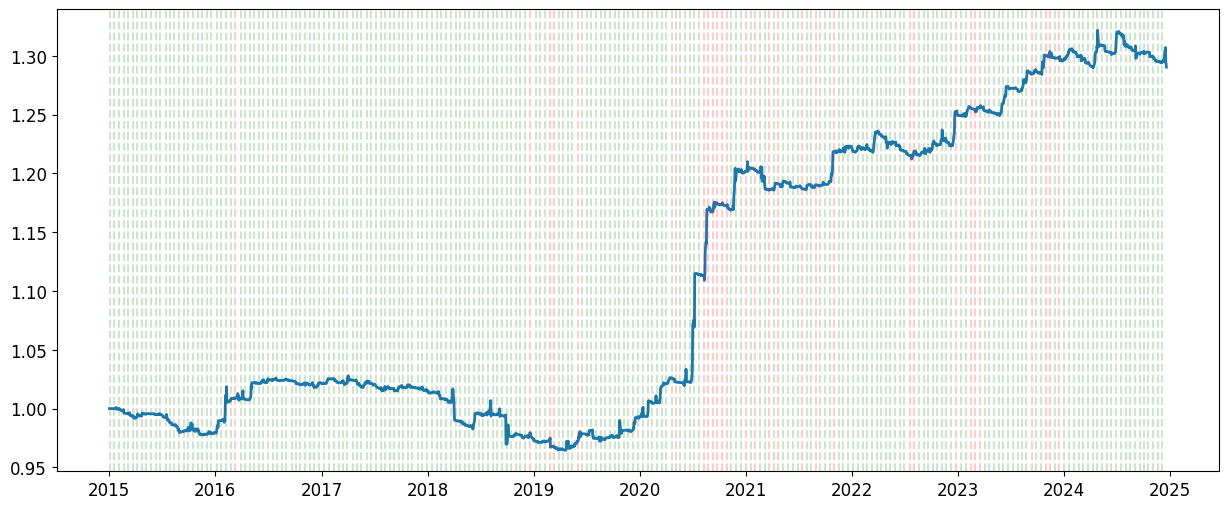}
%	\caption{Out-of-Sample Trading Performance with TSLA from 2015 to 2025. Rolling Optimization with Window Size 10 Days.}
\label{fig:tslatradingresultwinsize10}

\centering
\includegraphics[width=0.905\linewidth]{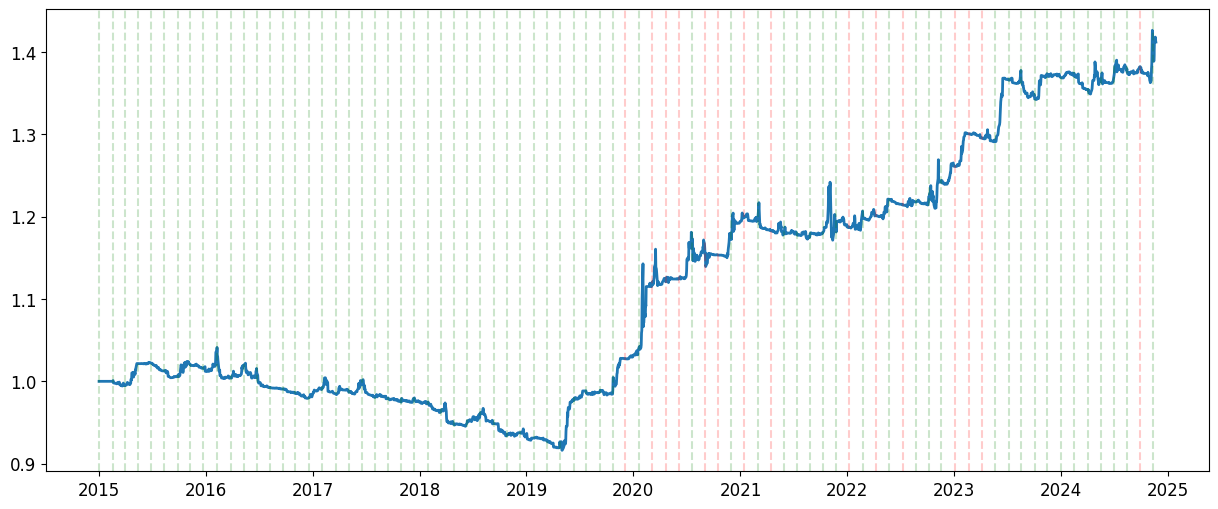}
%	\caption{Out-of-Sample Trading Performance with TSLA from 2015 to 2025. Rolling Optimization with Window Size 30 Days.}
\label{fig:tslatradingresultwinsize30}

\centering
\hspace{-.1cm}\includegraphics[width=0.91\linewidth]{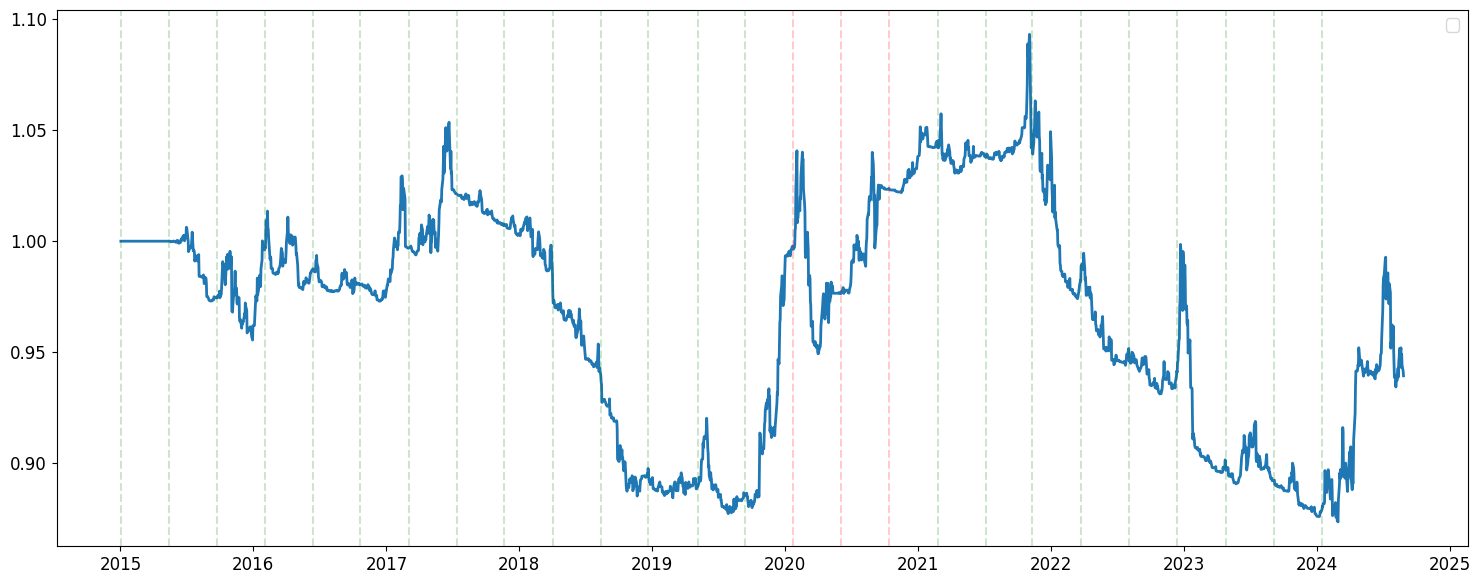}
\caption{Out-of-Sample Trading Performance with TSLA from~2015 to 2025 Using Rolling Optimization with Window Size~$N=10, 30, 90$ Days.}
\label{fig:tslatradingresultwinsize90}
\end{figure}

The time-varying optimal triple is depicted in the lower panel of Figure~\ref{fig: tslatradingresults}, while the trading performance, in terms of the account value~$V(k)$, is shown in the upper panel.
The solid blue line in the figure represents the cumulative gain using the actual TSLA price data. 
Additionally, green dotted vertical lines correspond to instances where the balanced policy was applied, while red dotted vertical lines represent the application of the complementary policy. 
Notably, with a window size $m=60$, the portfolio value exhibits a clear upward trend in recent years, yielding an approximate~$6\%$ positive cumulative gain by the end of the trading period.  If a different window size is used, such as~$m=10,30,90$, as selected previously, the corresponding out-of-sample trading performance is shown in Figure~\ref{fig:tslatradingresultwinsize90}. In this specfic case, we see that~$m=10$ gives the best performance, yielding an approximately cumulative return of~$30\%$ with a drawdown of about $5\%$.

\begin{remark} \rm 
$(i)$ Although the returns in real financial data are not strictly independent, the framework demonstrates robust performance in practice. This observation suggests that while our theoretical analysis assumes independence, the proposed double linear policy is robust to mild dependence, further supporting the practical applicability of the method. 
$(ii)$ While the current empirical study focuses on performance without explicitly modeling transaction costs, the robustness of the proposed framework under nonzero transaction costs has been analyzed in~\cite{hsieh2022robustness}. Specifically, it was shown that when the market exhibits a clear trend, the proposed double linear policy remains effective even with transaction fees.
\end{remark}

\section{Conclusion} \label{SECTION: Conclusion}
In this paper, we proposed a robust framework for the optimal gain selection problem within the double linear policy framework, parameterized by the triple~$(\alpha, K_L, K_S)$. By integrating semi-infinite robust positive expectation (RPE) constraints and extending classical mean-variance optimization, the framework ensures consistent performance across various market conditions, including adverse or volatile environments. Unlike methods that rely on boundary checks or sample approximations, our approach transforms a computationally complex semi-infinite problem into a tractable one, without losing robustness across the entire uncertainty set.

We established two key structured polices: the Balanced Policy, $(\alpha, K_L, K_S) = (\tfrac{1}{2}, K, K)$ for $K \in \mathcal{K}$ and the Complementary Policy, $(\alpha, K_L, K_S) = (\alpha, 1-\alpha, \alpha)$ for~$\alpha \in (0,1)$. The existence and uniqueness of the optimal solution were shown to hold only under these structures.  
The results rely on tools from optimization theory, including Berge's Maximum and Minimum Theorems and the Extended Extreme Value Theorem, to ensure existence and uniqueness of robust optimal policies. To facilitate implementation, we proposed a graphical approach leveraging monotonicity properties, enabling efficient identification of optimal gains. Numerical examples validated the framework's robustness and practical effectiveness.

Our results open the door to advanced robust positivity constraints in a class of discrete-time stochastic systems. Unlike typical $\mathcal{H}_\infty$ or LMI-based designs, we derive an explicit structural classification of feedback gains that guarantee positivity in the presence of uncertainty.

A natural extension of the work is to consider dependent returns, such as {\rm serially correlated} or stationary returns.  
For example,   modeling returns $\{X(k): k \geq 0\}$ using time-series or regression factor models could provide deeper insights into real-world trading behavior. A typical example is the model proposed in~\cite{fama1996multifactor} where
$
X(k) = a(k) + bf(k) + e(k)
$
with~$a(k)$ is the intercept term,~$f(k)$ is the time-varying {\em factor} and~$e(k)$ is the error term. 
Exploring the RPE property within such models could provide valuable insights in practical trading scenarios. 
However, this extension presents a significant theoretical challenge, as the proof techniques for establishing RPE fundamentally rely on the independence assumption. Specifically, the key argument in Lemma~\ref{lemma: expected cumulative gain or loss and variance} and Theorem~\ref{theorem: Robust Positive Expectation}, which builds on the separability of expectations, would no longer hold under serial correlation.

Despite these theoretical challenges,  empirical results using real market data (Section~\ref{SECTION: Trading With Historical Data}) indicate that the double linear policy performs well,  even when the independence assumption is violated. This observation indicates that, while extending the theoretical guarantees to dependent returns remains an open question, the practical effectiveness of our approach may extend beyond the theoretical constraints of our current analysis.
This gap between theoretical guarantees and empirical performance highlights an important direction for future research: developing new mathematical frameworks that capture the robustness properties of trading strategies under more general return processes while preserving analytical tractability. Another possibility is to explore the serial dependence on lattice-based return sequence; as proposed in \cite{hsieh2023robust}, where a multi-asset portfolio cases are also studied.  Such an approach could offer a way to handle discrete return distributions while preserving robustness properties.

Another promising direction is to explore beyond idealized market assumptions by explicitly accounting for transaction costs. Initial results in \cite{hsieh2022robust} indicate that transaction costs can erode the robustness of positive expectation. Addressing this erosion while preserving RPE properties presents an important challenge.
One potential approach is to incorporate proportional transaction costs into the policy design and re-derive the RPE constraints accordingly.  Additionally, incorporating time-varying conditional volatility into the framework could enhance its applicability. Models such as ARCH or GARCH, see \cite{bollerslev1986generalized} are particularly suited for such an extension.

%\medskip
\appendix

\section{Technical Proofs} \label{Appendix: Technical Proofs} 
\counterwithin{theorem}{section}
This section collects several technical results in Sections~\ref{SECTION: Problem Formulation} and~\ref{SECTION: Robust Optimal Gain Selection}.

\subsection{Proof in Section~\ref{SECTION: Problem Formulation}} \label{Appendix: proofs in Problem Formulation}

\begin{proof}[Proof of Lemma~\ref{lemma: state positivity}]
For~$k=0,1,\dots,$ using Equation~\eqref{eq: account value eq} and observe that
\begin{align*}
V(k) 
%		&= \frac{V_0}{2}\left( {\prod\limits_{j = 0}^{k - 1} {\left( {1 + KX\left( j \right)} \right)}  + \prod\limits_{j = 0}^{k - 1} {\left( {1 - KX\left( j \right)} \right)} } \right) \\
&\geq {V_0}\left( \alpha( 1 + K_L X_{\min}  )^k  +  (1-\alpha)( 1 - K_S X_{\max}  )^k  \right)\\
&:=V_{\min}^*(k).
\end{align*}
Given that $K_i \in \mathcal{K}$ with $i \in \{L, S\}$, $\alpha \in [0,1]$ and $X_{\min} > -1$, it is readily verified that both of $1 + K_L X_{\min} \geq 0$ and $1 - K_S X_{\max} \geq 0$. 
Hence, for stage~$k \geq 1$,
$(1 + K_L X_{\min} )^k \geq 0$ and $(1 - K_S X_{\max} )^k \geq 0.$
It follows that $V_{\min}^*(k)\geq 0$ for all $k$ with probability one. Hence, 	we have $
\mathbb{P}(V(k) >0) = 1
$ for all $k$.
To complete the proof, it remains to show that the double linear policy is cash-financed. 
In particular,	fix $K_i \in \mathcal{K}$ with $i \in \{L, S\}$ and $k \geq 1$. We begin by recalling that $\pi(k)  = \pi_L(k) +\pi_S(k)$ where $\pi_L(k) = K_L V_L(k)$ and~$\pi_S(k) = -K_S V_S(k)$.
Since $K_L, K_S \in \mathcal{K}$, it is readily verified that  $V_L(k), V_S(k) \geq 0$ for all~$k\geq 1$ with probability one. 	
Using the triangle inequality and the fact that $V(k) = V_L(k) + V_S(k)$,  it follows that
\begin{align*}
|\pi(k)| 
\leq | \pi_L(k) | + | \pi_S(k)| 	&= K_L V_L(k) + K_S V_S(k) \\
&\leq K_{\max} V(k)
\leq V(k)
\end{align*}
where the last inequalities hold since $K \in \mathcal{K} = [0, K_{\max}]$ and $K_{\max} \leq 1$. 
\end{proof}

\subsection{Proofs in Section~\ref{SECTION: Robust Optimal Gain Selection}} \label{Appendix: proofs in Main Results}

\begin{proof}[Proof of Lemma~\ref{lemma: expected cumulative gain or loss and variance}]
The machinery for the proof is elementary.  In particular, let $k$ be fixed. Using the fact that the outcomes $X(k)$ are independent and $\mathbb{E}[X(k)] = \mu$ for all $k$, we have
\begin{align*}
& \mathbb{E}[ {\mathcal{G}}_k(\alpha, K_L, K_S; X)] \\
%		& = \mathbb{E}[\mathcal{G}(K,k)]\\
& = V_0 \mathbb{E}\left[ \alpha R_+(k)  + (1-\alpha) R_-(k)  - 1  \right]  \\
%	&= \frac{1}{2}\left( {\prod\limits_{j = 0}^{k - 1} {\left( {1 + K\mathbb{E}\left[ {X\left( j \right)} \right]} \right)}  + \prod\limits_{j = 0}^{k - 1} {\left( {1 - K\mathbb{E}\left[ {X\left( j \right)} \right]} \right)}  - 2} \right)V_0  \\
%	&= \frac{1}{2}\left( {\prod\limits_{j = 0}^{k - 1} {\left( {1 + K\mu } \right)}  + \prod\limits_{j = 0}^{k - 1} {\left( {1 - K\mu } \right)}  - 2} \right)V_0  \\
&= V_0 \left( \alpha \left( 1 + K_L \mu  \right)^k+ (1 - \alpha)\left( 1 - K_S \mu  \right)^k - 1 \right). 
\end{align*}

%
%\begin{lemma}\label{lemma: variance for g} For $k=0,1,\dots,N-1,$
%	The variance of the cumulative gain or loss is given by
%	\begin{align*}
%		{\rm var}(\mathcal{G}(K,k))
%			&= {\alpha^2 V_0^2}((1+K\mu)^2 + K^2 \sigma^2)^k \\
%			& +{(1-\alpha)^2 V_0^2} ( (1-K\mu)^2 +K^2\sigma^2)^k\\
%		& +{ 2\alpha (1-\alpha)V_0^2 }  (1-K^2(\sigma^2+ \mu^2))^k+V_0^2\\
%		& -2\alpha V_0^2 (1+K\mu)^k		-2(1-\alpha)V_0^2 (1-K\mu)^k\\
%		&-{V_0}^2 \left( \alpha \left( 1 + K\mu  \right)^k+ (1-\alpha)\left( 1 - K\mu  \right)^k - 1 \right)^2
%%		&=\frac{V_0^2}{4}  (1+2K\mu + K^2(\sigma^2+\mu^2))^k \\
%%		&+\frac{V_0^2}{4}  (1-2K\mu + K^2(\sigma^2+\mu^2))^k\\
%%		&+\frac{V_0^2}{2}     (1-K^2(\sigma^2+\mu^2))^k\\
%%		&-V_0^2  [(1+K\mu)^k +(1-K\mu)^k-1]\\
%%		&-\frac{V_0^2}{4}\left( {{{\left( {1 + K\mu } \right)}^k} + {{\left( {1 - K\mu } \right)}^k} - 2} \right)^2
%	\end{align*}
%	for $K \in \mathcal{K}$. Of course, the standard deviation of cumulative trading gain or loss is ${\rm std}(\mathcal{G}(K,k)):=\sqrt{{\rm var}(\mathcal{G}(K,k))}$.
%\end{lemma}
%
To obtain the closed-form expression for variance, we proceed with a proof by a lengthy but straightforward calculation. Specifically, we note that
\begin{align*}
& {\rm var}(\mathcal{G}_k(\alpha, K_L, K_S; X))\\
& =\mathbb{E}[\mathcal{G}_k^2(\alpha, K_L, K_S; X)] - (\mathbb{E}[\mathcal{G}_k(\alpha, K_L, K_S; X)])^2.
\end{align*}
Since the mean is obtained in the earlier part of the proof, it remains to calculate the second moment term in the equality above.  That is,
{\small	\begin{align*}
&	\mathbb{E} \left[ \mathcal{G}_k^2(\alpha, K_L, K_S; X) \right] \\
	&= {V_0^2} \mathbb{E}\left[  \left( \alpha R_+(k)  + (1-\alpha) R_-(k)  - 1 \right)^2 \right]\\
	&= {\alpha^2 V_0^2}\prod_{j = 0}^{k-1}\mathbb{E}\left[ ( 1 + K_L X(j) )^2 \right]\\
	&\quad +{(1-\alpha)^2 V_0^2} \prod_{j = 0}^{k-1}\mathbb{E}\left[ (1 - K_S X(j))^2\right]\\
	&\quad + { 2\alpha (1-\alpha)V_0^2 }    \prod_{j = 0}^{k-1}\mathbb{E}\left[(1+ K_L X(j))(1 -  K_S X(j))\right]\\
	&\quad -2\alpha V_0^2 \prod_{ j = 0 }^{k-1}\mathbb{E}\left[ 1 + K_L X(j) \right]\\
	&\quad -2(1-\alpha)V_0^2 \prod_{j = 0}^{k-1}\mathbb{E}\left[1 - K_SX(j)\right]+V_0^2.
\end{align*}
%	Now, with the aid of the facts that $X(k)$ are IID,  $\mathbb{E}[X(k)]=\mu$, and $\mathbb{E}[X^2(k)]=\sigma^2 + \mu^2$, we have	
}Given that~$X(k)$ are independent with $\mathbb{E}[X(k)] = \mu$ and $\mathbb{E}[X^2(k)] = \sigma^2 + \mu^2$ for all $k$,  applying the linearity of expected value, a lengthy but straightforward calculation, leads to
\begin{align*}
& {\rm var}(\mathcal{G}_k (\alpha, K_L, K_S; X))\\
&= {\alpha^2 V_0^2}((1 + K_L\mu)^2 + K_L^2 \sigma^2)^k  \\
&+{(1-\alpha)^2 V_0^2} ( (1 - K_S \mu)^2 + K_S^2\sigma^2)^k\\
&\quad +{ 2\alpha (1-\alpha)V_0^2 }  (1 + \mu (K_L- K_S) -K_L K_S (\sigma^2+ \mu^2))^k\\
&\quad -2\alpha V_0^2 (1+K_L\mu)^k-2(1-\alpha)V_0^2 (1-K_S\mu)^k+V_0^2\\
&\quad -{V_0}^2 \left( \alpha \left( 1 + K_L \mu  \right)^k+ (1-\alpha)\left( 1 - K_S \mu  \right)^k - 1 \right)^2.
%		&=\frac{V_0^2}{4}  (1+2K\mu + K^2(\sigma^2+\mu^2))^k \\
%		&\hspace{10mm}+\frac{V_0^2}{2} \left(    (1-K^2(\sigma^2+\mu^2))^k\right)\\
%		&\hspace{10mm}-V_0^2  (1+K\mu)^k\\
%		&\hspace{10mm}+\frac{V_0^2}{4}  (1-2K\mu + K^2(\sigma^2+\mu^2))^k\\
%		&\hspace{10mm}-V_0^2 \left((1-K\mu)^k-1\right)\\
%		&\hspace{10mm} -\frac{V_0^2}{4}\left( {{{\left( {1 + K\mu } \right)}^k} + {{\left( {1 - K\mu } \right)}^k} - 2} \right)^2.
\end{align*}
To complete the proof, we note that the standard deviation ${\rm std}(\mathcal{G}_k(\alpha, K_L, K_S; X))=\sqrt{{\rm var}(\mathcal{G}_k(\alpha, K_L, K_S; X))}$.
\end{proof}

%
%\begin{lemma} \label{lemma: auxilary positivity}
%For $x,y \in [-1,1]$, the inequality $(1+x)^k + (1-y)^k \geq 2$ for all $k \geq 1$ and with equality  if and only if $x= y.$
%\end{lemma}
%
%
%
%
%\begin{proof} [Proof of Lemma~\ref{lemma: auxilary positivity}]
%Assume $(1+x)^k + (1-y)^k \geq 2$ for all~$k$ and~$x,y \in [-1, 1]$. We aim to show $x=y$. Since the inequality holds for all~$k$, it holds for $k=1$. For this particular case, we have $(1+x) +(1-y)\geq 2$, which implies $x - y \geq 0$. Or equivalently, $x \geq y$. Now consider the inequality again at $k=1$ but assume the terms are reversed, that is $(1-x) + (1+y) \geq 2$. This implies $x - y\leq 0$. In combination with both results, we find $x=y.$
%
%On the other hand, suppose that $x = y$, then we obtain
%\[
%(1+x)^k + (1-y)^k = (1+x)^k + (1-x)^k. 
%\]
%By the virtue of the fact that $(1+x)^k + (1-x)^k \geq 2 $ for all $x \in [-1, 1]$ and $k \geq 1$ with equality if and only if $x=0$ or $k=1$. Hence, the proof is complete.
%\end{proof}

\begin{proof}[Proof of Lemma~\ref{corollary: Robust Expected Growth}] 
Let $k \geq 1$. With $\alpha = \frac{1}{2}$, Lemma~\ref{lemma: expected cumulative gain or loss and variance} tells us that
$
\mathbb{E}\left[ { \mathcal{G} }_k \left( \tfrac{1}{2}, K,K; X \right) \right] = \tfrac{V_0}{2} \left( f(K, k, \mu) - 2 \right)
$
where 
$
f_k(K, \mu) :=  (1+K\mu)^k + (1-K\mu)^k.
$
Then, to establish the desired robust growth property, it suffices to show that~$f_{k+1}(K, \mu) \geq f_k(K, \mu)$ for all $K \in \mathcal{K}$ and~$\mu > -1$. 
Observe that
\begin{align*}
f_{k+1}(K, \mu ) 
%	& =  (1+K\mu)^{k+1} + (1-K\mu)^{k+1} \\
& =  f_k( K, \mu ) + K\mu [ (1 + K\mu)^k  - (1 - K\mu)^k].
\end{align*}
Since $\mu \in (-1,  X_{\max}]$ and $K \in \mathcal{K}$, we have $ K\mu \in (-1, 1]$.
Hence,	if~$K\mu = 0$; i.e., $K=0$ or $\mu = 0$, it follows that~$f_{k+1}(K,  \mu)  = f_k(K, \mu)$.
On the other hand,	if $K\mu \neq 0$; i.e., $K \in \mathcal{K}$ with $K >0$ and $\mu \neq 0$, using the basic fact that $a ((1+a)^k - (1-a)^k) >0$ for $a \neq 0$ and all~$k\geq 1$, it follows that 
$
f_{k+1}(K, \mu) > f_k(K,  \mu).
$
Therefore, we conclude that $f_{k+1}(K,\mu) \geq f_k(K, \mu)$.
\end{proof}

\begin{proof}[Proof of Lemma~\ref{lemma: robust monotonicity for iid case}] 
Fix $k > 1$ and $\alpha = \tfrac{1}{2}$. Without loss of generality,  set $V_0:=1$.
To  show that~$\mathbb{E}[ {\mathcal{G}}_k(\frac{1}{2}, K,K; \mu)]$ is increasing, we note that
{\small   \begin{align*}
&\frac{\partial }{\partial K} \mathbb{E}[ {\mathcal{G}}_k(\tfrac{1}{2}, K, K; X)] = \frac{k \mu}{2} \left(  \left( 1 + K\mu  \right)^{k-1} - \left( 1 - K\mu  \right)^{k-1}  \right). 
\end{align*}
}Now we split into two cases:

{\em Case 1:} For $\mu \in [0, X_{\max}] $, since $k > 1$ and $K\in \mathcal{K}$, it is readily verified that $(1+K\mu) \geq (1-K\mu) \geq 0.$ Therefore,
% it suffices to show that 
$
\left( 1 + K\mu  \right)^{k-1} \geq \left( 1 - K\mu  \right)^{k-1} .
$
%Note that $K\mu \geq -K\mu$, hence $1+K\mu \geq 1-K\mu$. Taking $k-1$ power still preserves the sign of inequality. It follows that
%\[
%  \left( 1 + K\mu  \right)^{k-1} - \left( 1 - K\mu  \right)^{k-1} \geq 0.
%\] 
Hence, 
$\frac{\partial }{\partial K} \mathbb{E}[ {\mathcal{G}}_k(\frac{1}{2}, K, K; X)] \geq 0.$

{\em Case 2:} For $ \mu  \in (-1, 0)$, then $0\leq (1+K\mu) \leq ( 1-K\mu)$, which implies that $(1+K\mu)^{k-1} \leq ( 1-K\mu)^{k-1}$. 
Hence,  we again have
$
\frac{\partial }{\partial K} \mathbb{E}[ {\mathcal{G}}_k(\frac{1}{2}, K, K; X)]  \geq 0.
$

Combing the two cases above, we conclude that $\mathbb{E}[ {\mathcal{G}}_k(\frac{1}{2}, K, K; X)] $ is monotonically increasing in $K \in \mathcal{K}$. 
Additionally,  if $\mu \neq 0$ and~$k> 1$, it is readily verified that the derivative is strictly positive for all admissible $K>0$. 
In combination with the facts that~$\mathbb{E}[ {\mathcal{G}}_k(\frac{1}{2}, K, K; X)] $ is continuous on~$\mathcal{K}=[0, K_{\max}]$, differentiable on $(0, K_{\max})$, it implies that the expected cumulative gain function is strictly increasing on~$\mathcal{K}$; see also~\cite[Theorem~5.14]{apostol1974}.

Using a similar idea, we now prove that the variance increases monotonically in $K$.   Take the derivative of  variance, a straightforward calculation leads to
%\begin{align}  \label{variance derivative}
%	& \frac{\partial}{\partial K}{\rm var}(\mathcal{G}(K,k))\\ \notag
%	&=\frac{ k}{2}( (1+K\mu)^2 + K^2 \sigma^2 )^{ k-1 }( \mu + K (\sigma^2 + \mu^2) )\\ \notag
%	&+\frac{k}{2}( (1-K\mu)^2 +K^2 \sigma^2) )^{k-1}(-\mu +K(\sigma^2+\mu^2)) \\ \notag
%	&-{ k} (1-K^2(\sigma^2+\mu^2))^{k-1}K (\sigma^2 + \mu^2)\\ \notag
%	&-\frac{k \mu}{2} ((1+K\mu)^{k-1} -(1-K\mu)^{k-1} )[(1+K\mu)^k + (1-K\mu)^{k} ].
%\end{align}
{\small \begin{align*}   
&	\frac{\partial}{\partial K}{\rm var} \left( \mathcal{G}_k (\tfrac{1}{2}, K, K; X) \right)\\
&= \frac{ k\mu}{2}[( (1+K\mu)^2 + K^2 \sigma^2 )^{ k-1 } 	-( (1-K\mu)^2 +K^2 \sigma^2) )^{k-1} ] \\  
&+ \frac{ k}{2}( (1+K\mu)^2 + K^2 \sigma^2 )^{ k-1 } K (\sigma^2 + \mu^2) \\  
&+ \frac{k}{2}( (1-K\mu)^2 +K^2 \sigma^2) )^{k-1}K(\sigma^2+\mu^2) \\ 
&- { k} (1-K^2(\sigma^2+\mu^2))^{k-1}K (\sigma^2 + \mu^2)\\ \notag
&- \frac{k \mu}{2} ((1+K\mu)^{k-1} -(1-K\mu)^{k-1} )[(1+K\mu)^k + (1-K\mu)^{k} ].
\end{align*}
}Use  part~$(i)$ of Lemma~\ref{lemma: useful inequality for lower bound variance} to bound the second and third terms, we obtain
{\small \begin{align*}   
&	\frac{\partial}{\partial K}{\rm var}\left( \mathcal{G}_k(\tfrac{1}{2}, K, K; X) \right)\\
&\geq  \frac{ k\mu}{2}[( (1+K\mu)^2 + K^2 \sigma^2 )^{ k-1 } 	-( (1-K\mu)^2 +K^2 \sigma^2) ^{k-1} ] \\ 
&+ \frac{ k}{2}( (1+K\mu)^2  )^{ k-1 } 	+(( 1-K\mu)^2) )^{k-1} K (\sigma^2 + \mu^2) \\  
&- { k} (1-K^2(\sigma^2+\mu^2))^{k-1}K (\sigma^2 + \mu^2)\\ \notag
&- \frac{k \mu}{2} ((1+K\mu)^{k-1} -(1-K\mu)^{k-1} )[(1+K\mu)^k + (1-K\mu)^{k} ].
\end{align*}
}Now, observe that the last term
\begin{align*}
& ((1+K\mu)^{k-1} -(1-K\mu)^{k-1} )[(1+K\mu)^k + (1-K\mu)^{k} ] \\
%&\;\;\;\; = (1+K\mu)^{2k-1} -(1-K\mu)^{2k-1}   -2K\mu (1-K^2 \mu^2)^{k-1}  \\
& = ({(1+K\mu)^2})^{k-1}   - ({(1-K\mu)^2})^{k-1} \\
&\;\;\;\;\;\;\;+K\mu [ ( (1+K\mu)^2)^{k-1} + ((1-K\mu)^2)^{k-1} ]\\
&\;\;\;\;\;\;\; -2 K\mu(1-K^2 \mu^2)^{k-1}  .
\end{align*}
Thus, using the equality above, a lengthy but straightforward calculation leads to
{\small \begin{align}     \label{ineq: var_derivative 1}
&	\frac{\partial}{\partial K}{\rm var}\left(\mathcal{G}_k ( \tfrac{1}{2}, K, K; X) \right) \\
&\geq  	  \frac{ k\mu}{2} \left[( (1+K\mu)^2 + K^2 \sigma^2 )^{ k-1 } 	-( (1-K\mu)^2 +K^2 \sigma^2) ^{k-1} \right] \notag \\ 
&+ \frac{ k}{2}( (1+K\mu)^2  )^{ k-1 } 	+(( 1-K\mu)^2) )^{k-1} K \sigma^2  \notag \\  
&-  k (1-K^2(\sigma^2+\mu^2))^{k-1}K (\sigma^2 + \mu^2)\notag \\ 
&- \frac{k \mu}{2} [({(1+K\mu)^2})^{k-1}   - ({(1-K\mu)^2})^{k-1} ] \notag \\
&+{k K\mu^2} (1-K^2 \mu^2)^{k-1}.
\end{align}
}Applying part~$(ii)$ of Lemma~\ref{lemma: useful inequality for lower bound variance}; i.e., 
\begin{align*}  
& { \mu}[( (1+K\mu)^2 + K^2 \sigma^2 )^{ k-1 } 	- ( (1-K\mu)^2 +K^2 \sigma^2) ^{k-1} ] \\
& \geq  { \mu} [ ((1+K\mu)^2)^{k-1}   - ((1-K\mu)^2)^{k-1}]
\end{align*}
to lower bound the first term of inequality~\eqref{ineq: var_derivative 1}, we have
\begin{align}  \label{ineq: ineq of d_var}
&\frac{\partial}{\partial K}{\rm var} \left( \mathcal{G}_k (\tfrac{1}{2}, K, K; X) \right) \\
& \geq  	  \frac{ k}{2}( (1+K\mu)^2  )^{ k-1 } \\
&	+(( 1-K\mu)^2 )^{k-1} K \sigma^2  
- { k} (1-K^2(\sigma^2+\mu^2))^{k-1}K \sigma^2 \notag \\ 
&\;\;\; +{k K\mu^2} [(1-K^2 \mu^2)^{k-1}-  (1-K^2(\sigma^2+\mu^2))^{k-1}].
\end{align}
Note that since $K \in \mathcal{K}$, the term 
$$
1-K^2(\sigma^2+\mu^2) = 1 - K^2 \mathbb{E}[X^2(0)] \geq 1 - K^2 X_{\max}^2 \geq 0.
$$ 
Therefore, the last  term of inequality~\eqref{ineq: ineq of d_var} is nonnegative since $(1-K^2 \mu^2)^{k-1} \geq (1-K^2(\sigma^2+\mu^2))^{k-1}$. Also,
by part~$(iii)$ of Lemma~\ref{lemma: useful inequality for lower bound variance}, 
we have
$$
( (1+K\mu)^2 )^{ k-1 } 	+( (1-K\mu)^2  )^{k-1} \geq 2(1 - K^2\mu^2 )^{k-1}.
$$
This implies that the sum of the first two terms of inequality~\eqref{ineq: ineq of d_var}  is also nonnegative. 
%That is,
%\begin{align*}  
%	& \frac{\partial}{\partial K}{\rm var}(\mathcal{G}(K,k)) \notag\\ 
%	&\geq  	  { k}K \sigma^2[(1 + K^2\mu^2 )^{k-1}   - (1-K^2(\sigma^2+\mu^2))^{k-1}] \\ 
%	&+{k K\mu^2} [(1-K^2 \mu^2)^{k-1}-  (1-K^2(\sigma^2+\mu^2))^{k-1}].
%\end{align*}
Therefore, we conclude~$
\frac{\partial}{\partial K}{\rm var}(\mathcal{G}_k(\frac{1}{2}, K, K; X)) \geq 0
$
and the variance is monotonically increasing.

To complete the proof, we note that if~$\sigma>0$, it is readily verified that the lower bound~\eqref{ineq: ineq of d_var} is strictly positive for all admissible $K>0$. Hence, it follows that $\frac{\partial }{\partial K}{\rm var}({\mathcal{G}}_k(\frac{1}{2}, K, K; X)) > 0$.
In combination with the facts that ${\rm var}(\mathcal{G}_k(\frac{1}{2},K, K; X))$ is continuous on $\mathcal{K} = [ 0, K_{\max}]$, differentiable on $(0, K_{\max})$, it follows  that the variance is strictly increasing on $\mathcal{K}$.
\end{proof}

%\medskip
\begin{lemma}  \label{lemma: useful inequality for lower bound variance}
Consider the double linear linear policy with $(\alpha, K) \in \{\frac{1}{2}\} \times \mathcal{K}$.
Given $\mu > -1$, $\sigma^2 \geq 0$, and~$k\geq 2$, the following three inequalities holds. That is,\\
$(i)$
\begin{align} \label{ineq: useful ineq 2}
& ( (1+K\mu)^2 + K^2 \sigma^2 )^{ k-1 } 	+( (1-K\mu)^2 +K^2 \sigma^2) )^{k-1} \notag \\
&  \geq ( (1+K\mu)^2  )^{ k-1 } 	+(( 1-K\mu)^2) )^{k-1}
\end{align}
$(ii)$
\begin{align}  \label{ineq: useful ineq 1}
& \mu \left[ \left( (1+K\mu)^2 + K^2 \sigma^2 \right)^{ k-1 } - \left( (1-K\mu)^2 +K^2 \sigma^2 \right) ^{k-1} \right] \notag \\
& \geq   \mu [( (1+K\mu)^2)^{k-1}   - ( (1-K\mu)^2)^{k-1}]
\end{align}
and 
$(iii)$ 
\begin{align} \label{ineq: useful ineq 3}
( (1+K\mu)^2 )^{ k-1 } 	+( (1-K\mu)^2 ) )^{k-1}
& \geq 2(1 - K^2\mu^2 )^{k-1}.
\end{align}
%and $(iv)$
%\begin{align} \label{ineq: useful ineq 4}
% 1-K^2(\sigma^2+\mu^2) \geq 0.
%\end{align}
\end{lemma}

%\begin{proof}
%	See Appendix~\ref{Appendix: Technical Proofs}.
%\end{proof}

\begin{proof}[Proof of Lemma~\ref{lemma: useful inequality for lower bound variance}] 
%We note that the inequalities in parts~$(ii$) and $(iii)$ are symmetric in $\mu$. Hence, in the sequel, without loss of generality, we assume that $\mu \geq 0.$ 
Part~$(i)$ holds trivially by noting that $K^2 \sigma^2 \geq 0$.

For~$(ii)$, note that for $k=2$, the left-hand side of the stated inequality~\eqref{ineq: useful ineq 1} simplifies to
$
\mu [ (1+K\mu)^2  - (1-K\mu)^2   ] 
$ 
which satisfies the inequality automatically.
Now, we proceed a proof by induction. Assuming the stated inequality holds for~$k-1$, we must show that the inequality also holds for $k.$ Set a shorthand function 
$$
f_{\pm}(K, \mu, \sigma) :=  (1 \pm K\mu)^2 + K^2 \sigma^2 . 
$$
Observe that, for $\mu \geq 0,$ 
\begin{align*}
& \mu \left[ f_{+}(K, \mu, \sigma)^k -f_{-}(K, \mu, \sigma)^k \right] \\
&= \mu f_{+}(K, \mu, \sigma)^{k-1} \left( (1+K\mu)^2 + K^2 \sigma^2 \right) \\
&\qquad  -\mu f_{-}(K, \mu, \sigma)^{k-1}  \left( (1-K\mu)^2 +K^2 \sigma^2 \right)\\
&\geq \mu (1+K^2\mu^2)\left( ((1+K\mu)^2  )^{k-1} -( (1-K\mu)^2 )^{k-1} \right) \\
&\qquad +2K\mu^2 \left[ f_+(K,\mu,\sigma)^{k-1} +f_{-}(K, \mu, \sigma)^{k-1} \right].
\end{align*}
where the last inequality holds by applying the inductive hypothesis.
Now,  with the aid of part~$(i)$, we obtain
%{ 
%\begin{align*}
%& f_+(K,\mu,\sigma)^{k-1} +f_{-}(K, \mu, \sigma)^{k-1} \\
%& \;\;\;\; \geq ( (1+K\mu)^2  )^{ k-1 } 	+(( 1-K\mu)^2) )^{k-1}
%\end{align*}
%}which implies that
\begin{align*}
& \mu \left[ f_{+}(K, \mu, \sigma)^k -f_{-}(K, \mu, \sigma)^k \right] \\
&\geq \mu (1+K^2\mu^2) \left( ((1+K\mu)^2  )^{k-1} -( (1-K\mu)^2 )^{k-1} \right)\\
&\qquad +2K\mu^2 \left[ ( (1+K\mu)^2  )^{ k-1 } 	+(( 1-K\mu)^2) )^{k-1}\right]\\
& = \mu [( (1+K\mu)^2)^{k} -   ( (1-K\mu)^2 )^{k}]
\end{align*}
and  the proof of part~$(ii)$ is complete.

%To prove part~$(iii)$, we again proceed a proof by induction. Note that for $k=2,$ the left-hand side of inequality~(\ref{ineq: useful ineq 3}) becomes $(1+K\mu)^2 + (1-K\mu)^2 = 2(1+K^2\mu^2)$, which is no less than the  right-hand side; hence, the inequality is satisfied for $k=2$. Next, we assume the inequality~(\ref{ineq: useful ineq 3}) holds for $k-1$, and we must prove it holds for $k$. Observe that
%\begin{align*}
%	&( (1+K\mu)^2 )^ k  	+( (1-K\mu)^2 ) )^k \\
%	&=	( (1+K\mu)^2 )^{ k-1 } (1+K\mu)^2 	+( (1-K\mu)^2 ) )^{k-1} (1-K\mu)^2\\
%	&\geq 		(1+K^2\mu^2) 2(1 + K^2\mu^2 )^{k-1}. \\
%	&+2K\mu  [ ( (1+K\mu)^2 )^{ k-1 } - ( (1 - K\mu)^2 )^{ k-1 } ] \\
%\end{align*}
%where the last inequality hold because of the fact that $\mu \geq 0$, $K^2(\mu^2 + \sigma^2) $, and $	( (1+K\mu)^2 )^{ k-1 } (1+K\mu)^2 	- ( (1-K\mu)^2 ) )^{k-1}  \geq 0.$
To prove part~$(iii)$, 
%it suffices to show that
%\begin{align} \label{ineq: equiv}
%	( (1+K\mu)^2 )^{ k-1 } 	+( (1-K\mu)^2 ) )^{k-1}
%& - 2(1 - K^2\mu^2 )^{k-1} \geq 0
%\end{align}
we note that since $-1 < \mu \leq X_{\max}$ and $K \leq K_{\max}$, it is readily verified that $1-K^2\mu^2 \geq 0.$
Therefore, we observe that 
%write left-hand side of the inequality~(\ref{ineq: equiv}) above as follows:
{\small \begin{align*}
&	\left( (1+K\mu)^{ k-1 })^2 	+( (1-K\mu)^{k-1} \right)^2
- 2(1 - K \mu )^{k-1} (1+K\mu)^{k-1} \\
& = \left( (1+K\mu)^{k-1} - (1-K\mu)^{k-1} \right)^2 \geq 0. \hspace{2.5cm} \qedhere
\end{align*}
} \end{proof}

\bibliographystyle{apalike}
\bibliography{refs}

\end{document}